\documentclass[letterpaper,twocolumn,10pt]{article}

\usepackage{tikz}
\usepackage{graphicx}
\usepackage{amsfonts}
\usepackage{amsmath}
\usepackage{listings}
\usepackage{xcolor}
\usepackage{algorithm}
\usepackage[font=small]{caption}
\usepackage{subcaption}
\usepackage[noend]{algpseudocode}

\usepackage{usenix2019_v3}

\begin{document}
%-------------------------------------------------------------------------------

%don't want date printed
\date{}

\title{\Large \bf Biscuit: A Compiler Assisted Scheduler for Detecting and Mitigating Cache-Based Side Channel Attacks}

%for single author (just remove % characters)
\author{
{\rm Sharjeel Khan}\\
{Georgia Institute of Technology}\\
{smkhan@gatech.edu}
\and
{\rm Girish Mururu}\\
{Georgia Institute of Technology}\\
{girishmururu@gatech.edu}
\and
{\rm Santosh Pande}\\
{Georgia Institute of Technology}\\
{santosh@cc.gatech.edu}
} 

\maketitle

%-------------------------------------------------------------------------------
\begin{abstract}
%-------------------------------------------------------------------------------

Side channel attacks steal secret keys by cleverly leveraging information leakages and can, therefore, break encryption. Thus, detection and mitigation of side channel attacks is a very important problem, but the solutions proposed in the literature have limitations in that they do not work in a real-world multi-tenancy setting on servers, have high false positives, or have high overheads, thus limiting their applicability. In this work, we demonstrate a compiler guided scheduler, Biscuit, which detects with high accuracy cache-based side channel attacks for processes scheduled on multi-tenancy server farms. A key element of this solution involves the use of a cache-miss model which is inserted by the compiler at the entrances of loop nests to predict the cache misses of the corresponding loop. Such inserted library calls, or “beacons”, convey the cache miss information to the scheduler at run time, which uses it to co-schedule processes such that their combined cache footprint does not exceed the maximum capacity of the last level cache. The scheduled processes are then monitored for actual vs predicted cache misses, and when an anomaly is detected, the scheduler performs a search to isolate the attacker. We show that Biscuit is able to detect and mitigate Prime+Probe, Flush+Reload, and Flush+Flush attacks on OpenSSL cryptography algorithms with an F-score of 1, and also to detect and mitigate degradation of service on a vision application suite with an F-score of 0.9375. Under a no-attack scenario, the scheme poses low overheads (up to a maximum of 6\%). In the case of an attack, the scheme ends up with less than 11\% overhead and is able to reduce the degradation of service in some cases by 40\%. We believe that because of its many desirable and real-world features such as an ability to deal with multi-tenancy, its ability to detect attacks early, its ability to mitigate those attacks, and low runtime overheads, Biscuit is a practical solution.

\end{abstract}
%-------------------------------------------------------------------------------
\section{Introduction}
\label{sec:introduction}
%-------------------------------------------------------------------------------

% Introduce side channel attacks and danger

Modern servers are comprised of multi-core machines that run multiple processes in parallel.
These processes are isolated and protected from one another due to a separation of 
virtual address spaces which have different access permissions; in addition many servers also adopt virtual machines for multi-tenancy
to achieve complete software stack isolation. The purpose of isolation is to make 
the memory contents or private data of one process non-accessible to
other processes. In spite of such mechanisms there have been attempts to gain access to private data. 
Private data has been leaked or attacked traditionally by exploiting
memory corruption or deviating control flow of the processes through input 
strings~\cite{rop,jop}, for which strong defense mechanisms
~\cite{cfi, practical_cfi, control_flow_locking,
kcofi, opaque_cfi, uCFI, DBLP:conf/asplos/GeCJ17, 7920853}
have been proposed. While these mechanisms safeguard against the faults in the program 
itself, side channel attacks that do not rely on return oriented or jump oriented programming (ROP or JOP)
~\cite{rowhammer,spectre,meltdown,primeprobe,flushreload,flushflush}
are becoming ubiquitous.
Side channel attacks are a class of attacks that extract the
secret key in cryptography algorithms by recording the changes (or differential behaviors) in the
physical properties of the machine (which act as a 
side channel).
%The secret key is obtained
%by analyzing the differential information exhibited by the physical
%property which is used as a side channel.
Attacks have used different physical properties
of systems such as time~\cite{flushreload, primeprobe, flushflush}, power consumption
~\cite{power}, memory consumption~\cite{memory}, sound~\cite{acoustic}
or electromagnetic emissions~\cite{electromagnetic} to leak data.
   
Among various side channel attacks that leverage different physical properties,
time-based attacks that utilize caches are most prevalent and are attractive for the attackers
for the following reasons:
\begin{itemize}
    \item Caches are a major component of the data access pipeline and are used
           for reducing memory latency in
           all computer systems. Since caches are omni-present, such attacks can
           be staged on a wide variety of systems. 
    \item Cache-based side channel attacks are easier to perform esp. the monitoring
          phase of the attack is easier, since
          external equipment is not required. They can be carried out
          remotely without physically accessing the machine (which is a
          requirement in many other side channel attacks such as
          electromagnetic emissions).
    \item The attack also slows down the victim, thus degrading the
          performance of the service provided by the process to its users.
    %\item The attack also slows down the victim, which can be desirable
    %      because it degrades the
    %      performance of the service provided by the process to its users.
\end{itemize}   

Currently, the literature describes three well-known cache-based
side channel attack techniques that retrieve the cryptography key:
(i) Flush+Flush~\cite{flushflush},
(ii) Prime+Probe~\cite{primeprobe}, and (iii) Flush+Reload~\cite{flushreload}.
These attacks force the victim's data out of the cache  
and then record the pattern in which the cache sets or cache lines were
filled in by the victim. The adversary repeats this step and then
analyzes the cache access pattern to obtain the victim's 
secret key, as successfully demonstrated in 
% , which has been demonstrated successfully in
~\cite{tromer2010efficient, osvik2006cache, flushreload, yarom2014recovering}.
Note that even if the attacker is not able to decipher 
the secret key by analyzing the cache access pattern,
% to successful in terms of analyzing the 
% cache access pattern and retrieve the secret key,
she would have still managed to slow 
down the victim substantially thus degrading the service of the victim.
Degradation
of Service (DS) attack is often one of the side effects of the
key-stealing attack. 

% % Explain time-based attacks
% Out of all these side channel attacks, the time-based attacks are easier to monitor because there is no need for any
% external equipment to detect the physical property and the attack can be done and detected both locally and remotely
% \cite{remote}. The three important time-based cache attacks are Flush+Flush \cite{flushflush}, Prime+Probe \cite{primeprobe}, and Flush+Reload \cite{flushreload}. These time-based attacks remove the data from the cache then they check if the data is back in the cache based on the time it took to fetch the data. By continuously repeating these actions, they can extract  They have been used to obtain the secret key from cryptography algorithms \cite{tromer2010efficient, osvik2006cache, flushreload, yarom2014recovering}, keystrokes \cite{ristenpart2009hey}, or confidential system information \cite{meltdown}. 

\subsection{Cache-based side channel Attack Techniques}
Several cache-based side channel attacks (CSAs) have been studied in the literature.
Here, we focus on
three well-known and recent CSA
techniques, commonly referred to as cache attack techniques (CATs), that recover the
secret key from a cryptography algorithm.
\begin{enumerate} 
    \item \textbf{Flush+Reload:} This technique relies on identical
    cryptography code or data pages to be shared
    between the attacker and victim processes.
    The adversary
    also relies on a selected set of cache lines to be flushed out
    through the invocation of certain instructions, e.g. the
    \textit{clflush} instruction in X86.
    % The adversary conducts the attack in three steps.
    First, the attacker
    flushes a memory line from the cache. Then the attacker waits
    for a fixed interval during which the victim may access
    the memory line, which will cause the memory line to be fetched back into the cache. After
    the wait, the attacker accesses the memory line that was flushed. If the victim
    accessed the memory line and cached it, the attacker's access duration will
    be much shorter compared to accessing the line from the memory. 
    By continuously repeating the
    above steps, the attacker records the memory access pattern
    by the victim, which is later analyzed to deduce the secret key.  
    
    \item \textbf{Flush+Flush:} This technique is similar to the above
    Flush+Reload attack. It shares the 
    same requirements and the same first step.
    The technique leverages the fact that
    the flush instruction (e.g. \textit{clflush}) aborts early
    when the memory line is 
    not in the cache. By exploiting this fact,
    the attacker, rather than accessing
    the memory line after the first flush (as in Flush+Reload), flushes 
    the memory line again. If the victim did not access the 
    memory line in the intervening period between the first
    and second flush, then the attacker's second flush will abort early.
    If the victim did access it, however, then the second flush evicts the memory
    line from all the caches, which takes more time.
    % On the other 
    % hand, if the victim never accessed the line in the wait period, the
    % attacker's second flush will abort early in a shorter time. 
    The attacker records all the cache lines accessed by the victim and
    then analyzes the differential behavior to crack the secret key as
    in Flush+Reload. The second flush not only checks if a memory line
    was accessed, but also sets up the cache to check if the line is
    accessed again, thus eliminating an extra step.
    
    \item \textbf{Prime+Probe:} This technique does not have any
    setup requirements and hence can be much more pervasive.
    Before the attack is initiated, the attacker creates an eviction
    set, which is a set of known memory lines that will collide
    with a victim's cache set. 
    In the prime step, the attacker fills the entire cache
    set with the memory lines from the corresponding eviction set.
    % In the second step, similar to Flush and Reload,
    The attacker waits for a fixed interval of time, during which
    the victim may access a memory line from the cache set
    (thereby evicting a line from the eviction set).
    Then in the probe step, the attacker
    accesses the eviction set again, checking if any memory line
    in the eviction set has been removed from the cache by the victim.
    If the victim never evicted a cache line from the same cache set,
    then the accesses for each memory line in the eviction
    set will be short (and vice versa). The attacker records the cache
    access pattern to later analyze the secret key as in the above
    techniques. Similar to flush step in Flush+Flush, the probe 
    step not only checks if any line in the eviction set was evicted, but
    also sets up the cache for the next probe by filling cache lines
    with the eviction set.
    
\end{enumerate}
Regardless of the technique,
the above attacks cause heavy cache misses for the victim process.
The cache misses therefore can serve as the 
basis for detecting the attack. The key questions to be answered however
are: What is the expected behavior of normal cache
misses at a given program point during the application's dynamic
execution (a no-attack scenario), and how can one carefully modulate
the expected cache behaviors
such that the departures from the same are successfully declared as
attacks? Through a combination of compiler analyses, cache
models, and careful scheduling decisions, this work
successfully constructs such a solution. Before we delve into the details
of our scheme, we provide a detailed survey of current
solutions discussing their limitations. 
% Organize it as prevention and detection and mitigation

%In detection, talk about its mitigation mechanisms
\subsection{Defense Mechanisms}
The CATs shown so far directly target
the secret key in the cryptography  algorithms, and 
several works have tried to either prevent or to
detect and mitigate this.
% calls for a serious %(change word)
% look into possible defense mechanisms for these attacks.

\subsubsection{Prevention Techniques}
Several CAT prevention mechanisms focus on
changing the cache designs, such as changing the cache replacement policy~\cite{sharp}, 
encrypting the cache address~\cite{qureshi2018ceaser, encryptedcache},
or locking cache lines~\cite{wang2007new}.
These solutions require changes to the hardware
and hence do not apply to the already existing systems. In some cases,
the solutions degrade the performance of the applications. 
Software-based hard isolation prevents the sharing of resources that contain sensitive data. 

Cachebar~\cite{softwareapproach} is a memory management subsystem that
provides two main mechanisms against CATs.
The first prevents processes from sharing sensitive cache lines
by making private copies of shared pages.
A process using shared library pages containing cryptography
functions cannot be traced by an adversary using Flush+Reload
or Flush+Flush CAT. The second mechanism limits the number of
cache lines that a process can access, which inhibits the
Prime+Probe CAT from exercising the entire eviction sets.
The attacker is not able to retrieve all the accesses of the victim.
However, these mechanisms are closely tied to the workings of
specific CATs. Further, creating private, duplicate pages not only adds
performance overheads but also sheds the benefits of shared libraries.
Limiting the cache access adversely impacts the performance of
genuine processes with overheads up to 25\%.  

StealthMem~\cite{10.5555/2362793.2362804} allocates isolated
pages called ``stealth pages'' to each process,
which map to unique cache sets
(i.e. no other page can map to them).
StealthMem assumes confidential data and calculations
are placed within these stealth pages. To adhere to this constraint,
the source code of the sensitive processes must be modified.
The stealth pages create a partition of the cache. For four cores with
a common last-level cache (LLC), the shared cache size reduces by 3\%.
The lost shared cache space increases with the number of cores that
share the cache. The overheads reported on 4 cores with 6 VMs is at worst 11\%.
Modern machines comparatively have much higher core counts but do not have
a correspondingly large a cache, which can significantly increase the overhead
of StealthMem systems with 32 or 64 cores.
%The burden of rewriting applications plus the non-scalability
%of the design, are two key limitations of StealthMem that make it impracticable. 

While the previous approaches are a form of hard isolation, a
scheduler-based approach \cite{schedulerdefense} provides
a soft isolation software solution
that disrupts the recording of victims' cache access patterns by
pre-empting other processes. The work analyzes the minimum run time
guarantee and schedules the process with the corresponding time slices
to avoid adversaries from recording the cache accesses. The scheduler
also performs CPU state cleansing between pre-emptions, in order to create
a soft isolation between processes. This technique, however, increases
the latency of each process, and in server farm environments that
over-provision cores~\cite{overprovision},
frequently de-scheduling the processes and idling the
machines further decreases machine utilization and slows down execution.
Thus, both hard and soft isolation approaches have limitations that
can prevent adoption in practical settings.

\begin{figure}
    \centering
    \includegraphics[width=0.46\textwidth]{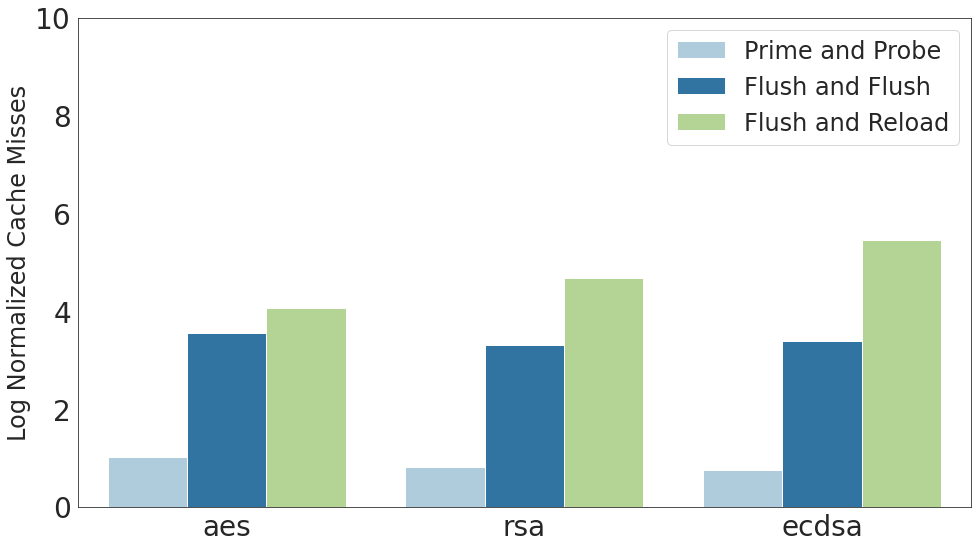}
    \caption{Log Normalized Cache Misses of Attacks}
    \label{fig:motivation}
\end{figure}

\subsubsection{Detection and Mitigation Techniques}
% Talk about detection of side channel attacks
There has been substantial research on detection and mitigation.
% Most of the detection mechanisms are software-based solutions
~\cite{236338, 203878} perform program
analysis on binaries to model the secret key-dependent
memory accesses and control flow.
The model is passed to an SMT solver to detect leakage areas 
that can be exploited by side channel attacks.
Program transformations ensure CPU cycles
and cache misses are independent of the secret data,
disrupting the timing channels~\cite{wu2018eliminating}.
% The first mitigation is replacing conditional
% statements with unconditional statements and inserting a new intrinsic
% function call CTSEL(c,t,f) which returns t if c is true and f if c is false.
% The second mitigation is to make memory accesses have the same timing whether
% a cache miss or hit. This is another mitigation technique but the entire analysis
% is done only for secret data like cryptographic keys.
However, these transformations result in longer response and
throughput times, with an average overhead of 50\% and a
worst case of 225\%.

Several techniques involve runtime mechanisms
~\cite{Kulah2019, zhang2016cloudradar, 8587756, 10.1016/j.asoc.2016.09.014}
that use performance counters to check for anomalies in
programs. Because of false positives in detecting anomalies, 
these runtime
detection mechanisms do not mitigate the attack but leave it
up to the system administrator for resolution.
In addition, these techniques are closely tied
to the cryptography algorithms for which they detect
the attack. For example,
SpyDetector~\cite{Kulah2019} is a semi-supervised
anomaly detection mechanism
to detect side channel attacks at runtime.  
The detection mechanism builds a clustering model that learns on 
cache misses, cache accesses, and the number of processes in
the execution windows. The predicted workload level is passed
to a clustering model, which raises an alarm for a possible
attack if a window is not within the cluster.
The clustering model is closely tied to the cryptography algorithms,
thus weakening the detection efficacy for algorithms with modifications
and for different workloads. Moreover, the mechanism must
figure out the granularity of the window that captures application
phases, because different applications
require different window sizes. This further reduces the generality of the mechanism. SpyDetector has a relatively high false positive and negative
rate, with false positive rates of up to 30\% and an F-score of
0.83 on Prime+Probe and Flush+Flush attacks.
% SpyDetector suffers from the burden of matching
% the execution windows of the victim and attacker, and
% finer window granularity incurs false positive rates
% of up to 30\%.

CloudRadar~\cite{zhang2016cloudradar} includes a
mitigation component. It generates a cache-access
profile of the cryptography applications with CATs, and
during runtime it flags behaviors that match
the profile as potential attacks.
After a match, CloudRadar also migrates one of the processes or 
a known victim process to mitigate the attack.
The execution profiles of the cryptography algorithms and the
cache-hit and cache-miss profiles of CATs can be noisy, leading
to many false anomalies.
As with SpyDetector,
different attack profiles and co-executing applications than
those seen in training can decreases the strength of this 
mechanism.
%A new, unknown attack with a slightly different profile
%or tricky modifications in the implementation of the
%known attacks decreases the strength of this 
%mechanism. Also, the behavior of these applications
%can change in the presence of other co-executing applications,
%which can lead to a profile mismatch. Attacks may escape the detection
%radar, or the mechanism may flag false anomalies.
Because
co-executing applications as well as variants of (known) attacks
are very likely in real execution environments, the defense
mechanisms must handle them.

In summary, most of the above mechanisms are either
hardware-based and do not apply to the existing machines,
or in the case of software-based solutions reduce the
efficiency of caches, increase latency, closely tie themselves
to the ``environment'' (i.e. the specific cryptography and cache
attack algorithms, the applications' performance profiles, etc.),
or do not perform well under multiple tenancy and due to the individual limitations discussed above
are not suitable to be adopted in a practical setting. 
%These mechanisms also fail to thwart the degradation of
%service effect by these CATs.
%Further, current hardware counter based runtime
%detection mechanisms 
%suffer from relatively high false positives and negatives.
% For example, SpyDetector has an
% F-score of 0.83 on Prime+Probe and Flush+Flush attacks.
% CloudRadar suffers from the burden of matching 
% the execution windows of the victim and attacker, and
% finer window granularity incurs false-positive rates
% of up to 30\%.

\subsubsection{Biscuit}
To overcome the above limitations, we propose Biscuit, a
compiler-assisted scheduler which in a multi-tenant 
environment detects cache-based side channel attacks on any program
and then mitigates the attack by de-scheduling the plausible culprit,
thereby avoiding any degradation of service attack.
The potential culprit is scheduled back and is allowed to run to
completion when all other processes have finished their execution.
%to enable continuity. 
As noted earlier, Biscuit relies on the fact that the victims of
CATs incur a significantly larger number of cache misses compared
to normal execution, as shown in Figure~\ref{fig:motivation}.
The number of cache misses is at least five times that of
the normal execution. In order to detect the cache
behavior anomaly, Biscuit must first
determine expected cache misses at certain program points
during the application's normal (no attack) execution, and it
must do this with significant accuracy. For this purpose, 
Biscuit first generates a cache-miss model for every loop
using compiler analysis. The model is then inserted before
the corresponding loop nest (a loop nest comprises of a sequence of several perfectly or imperfectly nested loops
at a give program point) in the application.
During the execution, this model transmits the predicted values of
cache misses of the loop nest to the scheduler.
{\it The predicted cache misses of a loop nest are proportional to the cache
footprint of the loop, because the predicted cache misses
are mainly due to the cold misses incurred by accessing unique memory references}.
This is so since the Biscuit's scheduler leverages these predicted cache footprints 
to schedule the processes such that the total of the cache footprints of all the
processes is less than  the last-level cache (LLC) capacity (or in other words, each processes cache footprint fully fits in
the LLC).
Since the cache footprints of scheduled processes fit in the cache,
under normal execution (no attack), the expected cache miss behavior
of each of the scheduled processes should remain unaltered (due to a
lack of any cache conflict; modern caches are highly associative
and thus mostly capacity conflicts occur in modern caches).
During execution, the scheduler monitors the cache misses
to check if the cache miss prediction for a loop is violated for a given process.
Upon encountering such a scenario, the scheduler carefully searches
for and isolates the culprit responsible for causing the victim to miss
heavily in the cache.

In short, 
Biscuit, by virtue of being the scheduler, can enforce (to a
reasonable degree, as we will see)
non-collisions in multi-tenancy. Because of this,
the departures from the predicted cache misses
must be attributable to some other reasons, viz. attacks.
This is a critical difference between Biscuit and other cache modeling techniques,
which tie themselves to the specific crypto-algorithms and/or suffer under
different co-execution environments; this results in the invalidations of other cache models
under different (co)-execution conditions leading to higher false positives or negatives. \emph{On the other hand, through effective scheduling, Biscuit enforces the conditions
under which the cache miss model was generated to remain valid under any runtime condition. As a result, Biscuit's cache
miss model functions well even under different, co-executing applications
and cache-based attacks.}

% In case of cache-based side channel attacks on cryptography 
% algorithms the misses are so high that it easily exceeds the predicted misses.
% The technique assumes the victim is compiled with the cache-miss model and the 
% rest of the process may or may not be compiled with the model.
% Our model makes sure we schedule processes by maximizing space and cores. 
We evaluated Biscuit on all of the above three
CATs on OpenSSL's implementation of AES, RSA, and ECSDA
cryptography algorithms in a multi-tenant environment.
Biscuit caught  all attacks on cryptography
algorithms with no false positives.
We also checked the usefulness of these techniques on the 
San Diego Computer Vision benchmarks to see if we could catch
degradation of service attacks on them. Biscuit caught all
attacks on these benchmarks, as well, though these caused a few
false positives. % We can provide F score details and comparison with cloud-radar and
We contribute the following in this paper:
\begin{itemize}
    \item A predictive cache-miss model for each loop based on loop
            bounds.
    \item A scheduler that leverages the the above model to determine 
           the expected cache-misses. The scheduler
            ensures that the dynamic cache behavior does not degrade under no attack condition,
            that the cache attacks are detected, and attackers are pinpointed
            and mitigated by de-scheduling.
    \item An evaluation of our technique using three well-known
            side channel attacks on OpenSSL's
            cryptography algorithms (to catch and defend against secret key leaks),
            and on computer vision applications (to test the effectiveness and defend against
             degradation of service attacks).
    \item A demonstration that we are able to catch these attacks
          in a multi-tenant environment with attack-agnostic models
          and overheads of less than 6\% under a no-attack scenario, less than 11\% under an attack scenario, and with a 40\% reduction of service degradation in one benchmark's case. 
        %   the performance of the framework is excellent. In particular, the overheads of the scheme are 6\% under no attack scenario 
\end{itemize}

% Outline
The remainder of the paper is structured as follows. In Section 2,
we provide an overview of the entire framework. In Section 3, we
explain the cache-miss model. In Section 4, we cover the beacon framework
that interacts with the scheduler. We
explain the Biscuit scheduler in Section 5 and provide a
detailed evaluation of the Biscuit framework in Section 6. In Section 7,
we present a discussion of some prior mechanisms against cache-based 
side channel attacks. We provide conclusions in Section 8.

% and the attacks that we can detect. In Section 3, we discuss   In Section 4, we describe the Beacon framework which provides the cache footprint information needed for detecting these attacks to the scheduler. In Section 5, we describe the Biscuit scheduler which schedules processes based on the cache miss footprint and detects the cache based side channel attacks to mitigate them. In Section 6, we present the results by running our system alongside the attacks on the cryptography algorithms and computer vision applications. In Section 7 we present prior work in this area and in section 8 provide the conclusions.

%-------------------------------------------------------------------------------
\section{Overview} \label{sec:overview}
%-------------------------------------------------------------------------------

To catch CSAs, Biscuit 
relies on the fact that the victim  experiences a significantly
larger number of cache misses while being attacked than
during normal execution.
To establish the expected number of cache misses during
normal execution, Biscuit first
builds a cache-miss model with the help of the compiler for every
loop in the application. The loops capture the cache behaviors well,
and a per loop cache model 
is able to distinguish cache behavior from one loop nest to another. 
The cache-miss model learns the cache misses
with respect to the loop properties over training input data sets. 
Every loop in the application is then associated with a
cache-miss model at its entry point. 
Generation and insertion of the cache model before the loop nests
forms the \textbf{compilation phase} of Biscuit. 
All applications must go through the compilation phase for
effective scheduling so that the dynamic cache contention is
minimal under normal (no attack) execution. There is
one exception for which we do not assume anything.
The attacker (i.e. the attacking process) may or may not
go through the compilation phase; if it does, it will have beacons associated
with it with its cache behavior trained under isolation (or no attack condition). If it does not,
the attack process will lack beacons. As can be seen later, under either condition, Biscuit
is able to detect it as a rogue process and isolate it.
Cache misses can also be calculated through analytical models
~\cite{analytical}, but these models can handle only affine
accesses and model simple caches. Further improvements in
the analytical model will provide an alternative model to the
current cache-miss model, which is based on machine learning and is shown
to be very accurate for Biscuit.

In the \textbf{runtime phase}, the cache-miss model infers
the cache misses of each loop before executing the loop.
This predicted information is
passed to the scheduler, which first schedules the processes
such that there is no cache-contention between the co-executing
processes. Then the scheduler monitors the cache misses while
the loops in the applications execute.
If the cache misses experienced by the loop are more than
the predicted cache misses, the scheduler performs a
search over all the executing processes to catch the plausible
culprit causing the cache misses. The 
culprit is then de-scheduled until all the executing processes
finish, after which it is scheduled back to ensure continuity on
false positives.

%Attack Model
\subsection{Threat Models and Goals}
Biscuit's scope and focus is on the time-based cache attacks.
Contrast this with HIDE~\cite{zhuang2004hide}, for example.
In HIDE, the attacker uses the cache to induce differential address information
on the address bus to leak the secret key. This attack
requires physical access to the machine to snoop on the address bus.
Our technique can catch the attacks shown in HIDE so long
as the hardware is unmodified, because these attacks also cause high cache misses in the victim. Caches have been used as a side channel to
leak data in other well-known attacks such as spectre~\cite{spectre},
meltdown~\cite{meltdown}, and others~\cite{van2018foreshadow, 191010},
but these attacks exploit some other architectural features such as
speculative execution. That is, they only use the cache as a
micro-architectural covert channel to transmit the data.
In particular, they do not target the victim's cache and hence
tackling them is outside the scope of this work.

Within a multi-tenant environment the CSAs
are tougher to discern just through cache misses,
because extra cache misses can result
from co-executing processes and not from an attack.
We assume a multi-tenant environment with a mix of various applications
such that they do not overload the system, and any process that needs
more than the available resources is rescheduled onto a different node.
This assumption is justifiable because
the server farms utilize less than 50\% of the server machine by 
over-provisioning~\cite{overprovision},
and the cluster scheduler can be invoked to reschedule the process onto
another node in the cluster.
The Biscuit scheduler uses the compiler-inserted cache-miss
information to efficiently schedule the co-executing process
onto the available cores such that the processes' combined
memory footprint fits the cache of the server machine.
We assume that the plausible victim
process is compiled with the Biscuit compiler. Other processes in the system
are also compiled with the Biscuit compiler for the Biscuit scheduler
to schedule them efficiently by averting cache contention.
The attacker may or may not be compiled with the Biscuit compiler.

We assume that the  LLVM compiler that instruments 
the application, the Biscuit Runtime, and the
underlying hardware and OS are not compromised.
We assume the trusted compiler inserts valid cache miss models, and
the process does not falsify the information provided to the
scheduler. A secure handshake based on a secret key mechanism
is commonly used in such scenarios.
The goal of the scheduler is to detect any
cache-based side channel attack (CSA)
on any process, find the plausible culprit process
and, de-schedule it.

%-------------------------------------------------------------------------------
\section{Cache Miss Model}
\label{sec:cachemissmodel}
%-------------------------------------------------------------------------------

We develop a linear-regression model to predict the
cache misses incurred by a loop.
For a given loop, with enough training inputs for different
loop iterations of the loop, the pattern of cache misses
is well co-related to the increase in the number of iterations. Thus  
for a given loop, cache misses are directly proportional to 
the number of loop iterations, i.e.
\begin{equation}\label{equation:LT1}
    CM \propto N \implies CM = \alpha * N
\end{equation}
where CM is the cache misses for a loop which takes N loop iterations, 
and $\alpha$ is a constant. In the case of nested loops, the cache 
misses depend on the number of iterations for each loop within the nest.
The cache misses for the loop nest are a function of the loop iterations
of the loops in the nest, i,e.
\begin{equation}\label{equation:LT2}
    CM = f(N_1, N_2, \dots, N_n)
\end{equation}
where CM is the cache misses for the loop nest, with $N_{j}$ being 
the number of loop iterations for each loop j in the loop nest. 

We normalize the loops by running the LLVM compiler's loop-simplify 
pass. The loop-simplify pass transforms loops so that the
loop induction variable begins at zero, increments by 1 on each
iteration, and breaks at some upper bound. The pass also
adjusts instructions inside the loop to maintain correctness.
Loop simplification normalizes for and while loops
that do not have data-dependent terminating conditions.
% Note, this is 
% not only for \textbf{for loops}. It also changes while loops that 
% are not data-dependent to fixed number of iterations loops. For 
For example, the loop in Listing~\ref{code:l1} is normalized
to Listing~\ref{code:l2}.
\begin{lstlisting}[xleftmargin=.1\textwidth,caption={Loop},label=code:l1]
for(i = 10; i < N; i+=2){
    a = b + i;
}
\end{lstlisting}
\begin{lstlisting}[xleftmargin=.1\textwidth,caption={Normalized Loop},label=code:l2]
for(i = 0; i < N/2 - 5; i++){
    a = b + (i + 5) * 2
}
\end{lstlisting}
In the normalized loops, the loop bound (upper) is equal to the number of
loop iterations. Thus, the number of cache misses in
Equation~\ref{equation:LT2} can be re-written
as a function of the loop bounds of each loop in the loop nest, i.e.
\begin{equation}\label{equation:LT3}
    CM = f(lb_1, lb_2, \dots, lb_n)
\end{equation}
where $lb_{j}$ = $N_{j}$ and $lb_j$ is the loop bound for the $j$ 
loop inside the loop nest. Each loop within the nest individually
contributes to the cache misses, so we can transform
Equation~\ref{equation:LT3} into 
\begin{equation}\label{equation:LT4}
    CM = f_1(lb_1) + f_2(lb_1, lb_2) + \dots + f_n(lb_1, lb_2, \dots, lb_n).
\end{equation}
Using the relation in Equation~\ref{equation:LT1},
Equation~\ref{equation:LT4} can be written as
\begin{equation}\label{equation:LT5}
    CM = \alpha_1*lb_1 + \alpha_2*lb_1*lb_2 + \dots + \alpha_n*lb_1*lb_2*\dots*lb_n + \alpha_0
\end{equation}
in which $\alpha_0$ is some constant.
Through linear regression, we get each coefficient $\alpha_k$ and the 
constant term $\alpha_0$ representing the y-intercept.

We record the cache misses for a loop by reading the performance counters
during the execution of the loop.
To record this data, our LLVM pass instruments the application with the
perf API to start monitoring
cache misses at the pre-header of each loop and stop monitoring at
the exit blocks of the loop.
To handle cases for which loop bounds cannot be extracted statically, we
profile the number of loop iterations and cache-misses
during the training runs.
The cache misses and the number of loop iterations 
are fed to the linear regression
implementation in scikit-learn~\cite{scikit-learn}
to learn a linear cache-miss regression model as in
Equation~\ref{equation:LT5} for each loop.

The linear model will only predict a single value for the
cache misses, but the cache-misses can non-deterministically
vary even for the same loop with the same number of iterations
when executed several times. The variance is different for
small and large loop bounds.
To accommodate for variation due to non-determinism,
we calculate the standard deviation over multiple runs for
each loop and then determine 
the maximum ratio of standard deviation to average cache misses, i.e.
\begin{equation}\label{eq:ratio}
   k = max(\frac{\sigma_l}{\overline{cm_l}}),  \forall \texttt{ loop } l
\end{equation}
where $\sigma_l$ is the standard deviation of each loop $l$, and
$\overline{cm_l}$ is the average cache misses of the corresponding loop.
This ratio is then appended to the predicted value in
Equation~\ref{equation:LT5} to
account for non-determinism as
\begin{equation}\label{equation:LT7}
    CM(U) = CM + k*CM
\end{equation}
where CM(U) is the upper bound on the cache misses.
Note that we do not need a lower
bound, because for the 
attack to be flagged, the cache misses must exceed CM(U);
any misses within the upper bound can be safely considered as
normal execution.
%Also note that there is no time component here.
%That is, a spike in cache misses is the trigger -- not a sustained
%or irregular spike over time. Thus, ``slow''
%attacks that operate over a longer time range are just
%as detectable.
The linear model (Equation~\ref{equation:LT5}), along with the upper
bound Equation~\ref{equation:LT7}, is instrumented before
the loop header in the LLVM intermediate code (IR).
During runtime, these equations are evaluated
and CM(U) is passed as a beacon to the scheduler.
% from Equation~\ref{equation:LT5} and
% $k$ is the maximum constant ratio of standard deviation to cache misses.

% Each loop will have its own constant ratio but we choose the maximum constant across all loops in the file to allow calculations to be general in our model. This equation will accommodate the non-determinism in program's cache misses.

% With the linear model's coefficients, the loop bounds and the constant 
% ratio, the beacon can predict the cache misses and pass them to the scheduler.

The cache misses CM(U) serves two purposes for 
the Biscuit scheduler. One is to 
check if a process is experiencing more cache misses
than it normally should. The second is to schedule 
the process with as minimal cache contention as possible,
which the scheduler does by using the cache-miss information 
as the cache/memory footprint of the loop. 
Using the cache misses same as memory footprint of
the loop is valid due to the reasons which were explained earlier in the section 1.2.3. 
%-------------------------------------------------------------------------------
\section{Beacons}
\label{sec:beacon}
%-------------------------------------------------------------------------------

The cache miss Equations~\ref{equation:LT5} and ~\ref{equation:LT7}
are inserted in the pre-header 
of the loop. These equations are evaluated at run time using the loop bound
values to predict the cache misses that will be incurred by the loop.
The predicted cache misses value is passed to the scheduler 
through a function call of a library called lib-beacons.
The library interfaces 
with the scheduler through shared memory in our 
implementation, but other secure communication channels can also
be employed.

These beacons are classified based on the precision of the
loop bound information for the loop.
%The imprecision of the data 
%emerges from the type of loops.
Many loops iterate a fixed 
number of times, determined by the loop bounds. These loop
bounds can be compile-time unknown but still loop invariant; or in other words, the loops exhibit fixed trip counts.
Other loops, however, terminate based on some data-dependent condition,
and the number of loop iterations for such loops cannot be 
resolved at the pre-header of the loop. It is also difficult to
calculate the loop iterations for loops with non-affine control
variables. Based on the precision
of the cache-misses information (calculated from the loop-bound info)
passed to the scheduler,
the beacons are classified as either Precise or Expected Beacons.

\begin{enumerate}
\item { \bf Precise Beacon - } 
the loop bounds are loop invariant and known at the
pre-header of the loop. Notice this can apply to
not just one loop but a nesting of loops. One such loop nest, in 
which the loop bound within each nest is a loop nest
invariant is rectangular loops as shown in Listing~\ref{code:l3}. 
\begin{lstlisting}[xleftmargin=0.1\textwidth,caption={Rectangular Loop},label=code:l3]
for(int i = 0; i <= N; ++i)
{
  a[i] = i+1;
  for(int j = 0; j < M; ++j)
     a[j] = j+1; 
}
\end{lstlisting}
The cache misses based on Equation~\ref{equation:LT5} for this loop are
\begin{equation}
 CM = \alpha_1*{M*N} + \alpha_2*N + \alpha_0.
\end{equation}
% Similarly, in triangular loops, as shown in Listing~\ref{code:l4},
% although the inner loop bound is dependent on the loop index
% and outer loop index variables, the 
% inner loop executes for one iteration to upper bound number
% of iterations, respectively, for every iteration of the
% outer loop, thus executing for a total of $N(N)/2$ iterations.
% \begin{lstlisting}[xleftmargin=0.1\textwidth,caption={Triangular Loop},label=code:l4]
% for(int i = 0; i <= N; ++i)
% {
%   a[i] = i+1;
%   for(int j = 0; j < i; ++j)
%      a[j] = a[j] + (j-i); 
% }
% \end{lstlisting}
% The cache misses based on Equation~\ref{equation:LT5} for this
% triangular loop are
% \begin{equation}
%  CM = \alpha_1*\frac{N^{2}}{2} + \alpha_2*N + \alpha_0.
% \end{equation}.

\item { \bf Expected Beacon - } 
the loop bounds are either 
data dependent or loop control variables that are non-affine.
For these loops, the loop bound is either not known before the
execution of the loop (as in a data dependent loop) or cannot be
calculated because
of the limitations of the compiler tools (as in a non-affine loop).
For example, the loop bound of Listing~\ref{code:l5} is
data dependent, and the number of loop iterations is only known
after the loop terminates.
\begin{lstlisting}[xleftmargin=0.1\textwidth,caption={Data Dependent Loop},label=code:l5]
while(a[i] == i)
{
  i += 1;
}
\end{lstlisting}
For such loops, we calculate expected cache misses by using the
average number of loop iterations collected during the 
training phase of the cache-miss model generation.
After plugging the expected value into
Equation~\ref{equation:LT5}, the cache misses for the loop is
given by
\begin{equation}
 CM =  \alpha_1 * E + \alpha_0,
\end{equation}
where E is the expected (average) loop bound.
If the loop nest consists of both precise and expected loops, then
the loop is classified by the type of outermost loop (either
expected or precise).

\end{enumerate}

Both precise and expected beacons are first inserted in
the loop nest pre-header. However, the parent function
of this loop nest can be called inside some other functions'
loop nest, which can result in multiple beacon calls to the 
scheduler. In such cases, we interprocedurally hoist the
beacon calls outside the caller function's outermost loop. 
However, due to such hoisting, the beacon loop bound
variable information required for precise beacons
may not be available at the external loop nest pre-headers
(since it may be interprocedurally defined inside the callee function).
In some cases, a backward interprocedural slice can be used to determine
the inner loop bound, but in other cases, estimates must be used. 
For the latter case, the precise beacons are converted to expected 
beacons, and the expected loop iterations are used instead at
the external loop nest pre-header.
In case of recursion, we detect the recursion cycle and the
incoming and outgoing edges of the cycle, and then hoist the beacon calls
on these edges appropriately. Hoisting for function pointers is limited
by the pointer analysis. In case of conditional calls, we hoist the
calls within the conditions -- except for conditions within the
loop. This is a tradeoff between approximate information
versus the overhead incurred by repeated beacons within the loop.
The LLVM beacon pass also inserts a completion beacon at each
exit of the loop nest. This completion beacon signals the
scheduler that the loop has ended. %to make scheduling decisions.
\section{Biscuit Scheduler}
\label{sec:scheduler}
%-------------------------------------------------------------------------------

The Biscuit scheduler uses the information sent by the beacon
for two purposes.
First, the Biscuit scheduler must ensure that the total
expected cache footprint of
all the co-executing applications does not exceed the LLC capacity. 
This ensures that the predicted cache misses must roughly equal
dynamic cache misses
for each scheduled application under normal (no attack) conditions.
% First, the scheduler uses the predicted cache misses
% to efficiently schedule the processes on the available cores, such that
% the cache contention is minimal among the simultaneously
% executing processes.
% When the cache misses are profiled with no other co-executing process,
% the total cache misses are roughly equal to the memory
% footprint of the executing
% process (or in this case the executing loop region).
Second, the scheduler uses the predicted
cache misses to detect a plausible attack when the
monitored cache misses of a loop exceeds the 
predicted cache misses. Because the scheduler avoids
cache contention, cache misses of the application 
will mostly exceed predictions because of a CSA. These two functionalities
of the Biscuit scheduler are detailed below.

% The cache misses is a good 
% approximation of the process's memory footprint because the misses
% provides the knowledge of how many cache spaces the process requires.

    %   \State {$\Call{resume\_process}{\texttt{descheduledProcs}}$}
        
    %     \ForAll{$\texttt{proc} \in \texttt{active procs - victim}$}
    %         % \If{$\texttt{proc} == \texttt{victim}$}
    %         %     \State {$\texttt{continue}$} 
    %         % \EndIf
    %         \State {$\Call{deschedule\_process}{proc}$}
    %         \State {$\texttt{c#processes} \gets \texttt{activeProcs.size} $}
    %         \If{$\texttt{c#processes} == \texttt{#processes}/2$}
    %             \State {$\texttt{break}$} 
    %         \Else
    %         \EndIf
    %     \EndFor
    %     \State {$\texttt{newMPKI} \gets \Call{getMPKI}{victim}$}
    %     \If{$\texttt{newMPKI} < \texttt{prevMPKI}$}
    %         \State {$\texttt{tempProcs} \gets \texttt{activeProcs}$}
    %         \State {$\texttt{activeProcs} \gets \texttt{[]}$}
    %         \ForAll{$\texttt{proc} \in \texttt{descheduledProcs}$}
    %             \State {$\Call{resume\_process}{proc}$}
    %         \EndFor
    %         \ForAll{$\texttt{proc} \in \texttt{tempProcs}$}
    %             \State {$\Call{deschedule\_process}{proc}$}
    %         \EndFor
    %         \State {$\texttt{descheduledProcs} \gets \texttt{tempProcs}$}
    %     \EndIf
    %   \EndWhile
    %   \State {$\Call{resume\_process}{descheduled processes - attacker}$}
      %\Comment { call returns a set of phi instructions }

\subsection{Scheduling}
Biscuit starts by scheduling a process on a unique core 
of every socket in the machine.
Biscuit knows the 
number of sockets, the number of cores per socket and the cache size of
each socket.
% thus scheduling every new process on a
% free core.
Whenever Biscuit receives a beacon from an executing process,
it checks if the memory
footprint (predicted cache misses) fits the available cache size. For 
the very first beacon, the available cache size is the cache size of
the socket the process is executing on. The available cache size is
updated by subtracting the cache requirements at every beacon, and
by adding back that amount at every loop-completion beacon.
% executing beacon of a scheduled process. 

On every beacon event, the processes are allowed to continue on the
socket only if the memory footprint (predicted cache misses)
of the beacon is less than or equal to the available cache.
If the memory footprint of the beacon exceeds the available cache,
the scheduler first checks if another socket with a free core can
satisfy the cache requirements. If not, then the scheduler
tries to greedily swap a process from another socket with
this process, such that the cache requirements of both the
processes is satisfied. If the 
scheduler cannot relocate the process, then the process is de-scheduled,
owing to a lack of adequate resources. At this point, the process
can be scheduled on another node in a cluster, if needed.
% Upon starting, the scheduler initializes its internal state and the 
% shared memory used to communicate with the beacon library.
% On executing a beacon at runtime, the predicted cache-misses info
% of the loop is written into the shared memory.
% The 
% scheduler reads a config file, containing system information and 
% process information. The system information is the number of sockets, 
% number of cores per socket, and the total cache size. On the other 
% hand, the process information is the process being scheduled plus 
% its inputs.
% The processes are scheduled based on their order in the 
% config file, i.e the process at the top of config file gets scheduled 
% first and the process at the end of the file gets scheduled last. 
% The scheduler makes sure to only schedule a process onto a core if 
% there is an empty core available or there is space in the cache for 
% the new process' misses. If there are processes which were not scheduled, 
% we keep them in a queue and schedule them whenever a process completes or 
% gets de-scheduled provided it fits within total cache size.

% For example, in figure~\ref{TODO}, we schedule one victim process 
% (RSA in our case), one other process, and one attack process (Flush 
% and Reload). The victim and attacker are scheduled on socket 1 
% while the the other process is scheduled on socket 2. By being scheduled on 
% the same socket, the attacker and victim has the same last-level cache. 

\begin{algorithm}
  \caption{Mitigation Algorithm}\label{Alg:mitigation}
  \begin{algorithmic}[1]
    \Procedure{Mitigate\_Attacker}{$victim$} 
    %   \State {$\texttt{descheduledProcs} \gets \texttt{[]} $}
        \While{$\texttt{\#suspectProcesses} > 8$}
        %  \State {$\texttt{\#processes} \gets \texttt{activeProcs.size} $}
            \State {$\texttt{prevMPKI} \gets \Call{getMPKI}{victim}$}
            \State {$\texttt{setA} \gets \newline \Call{deschedule\_half\_processes}{suspectProcesses}$}
            \State {$\triangleright \texttt{ Descheduled Processes is set A}$}
            
            \State {$\texttt{setB} \gets suspectProcesses - setA $}
            \State {$\triangleright \texttt{ Active Processes is set B}$}
            
            \State {$\texttt{newMPKI} \gets \Call{getMPKI}{victim}$}
            \If{$\texttt{newMPKI} < \texttt{prevMPKI}$}
                \State {$\Call{resume\_processes}{setA}$}
                \State {$\texttt{suspectProcesses} \gets setA$}
                \State {$\triangleright \texttt{ Attacker is in set A}$}
            \Else
                \State {$\Call{resume\_processes}{setA}$}
                \State {$\texttt{suspectProcesses} \gets setB$}
                \State {$\triangleright \texttt{ Attacker is in set B}$}
            \EndIf
        \EndWhile
        \For{$\texttt{proc in suspectProcesses}$}   
            \State {$\texttt{prevMPKI} \gets \Call{getMPKI}{victim}$}
            \State {$\Call{deschedule\_process}{proc}$}
            \State {$\texttt{newMPKI} \gets \Call{getMPKI}{victim}$}
            \If{$\texttt{newMPKI} < \texttt{prevMPKI}$}
                \State {$\texttt{attacker} \gets proc$}
                \State {$\triangleright \texttt{ Attacker is proc}$}
            \Else
                \State {$\Call{resume\_process}{proc}$} 
                \State {$\triangleright \texttt{ Attacker is not proc }$}
                \State {$\texttt{so re-schedule it}$}
            \EndIf
        \EndFor
        \If{$\texttt{attacker} == \texttt{None}$}
            \State {$\Call{deschedule\_process}{victim}$} 
            \State {$\texttt{attacker} \gets victim$}
            \State {$\triangleright \texttt{ No attacker found }$}
                \State {$\texttt{so victim is attacker}$}
        \EndIf
    \EndProcedure
  \end{algorithmic}
\end{algorithm}

\subsection{Detection and Mitigation of Side Channel Attacks}
Immediately after Biscuit has handled a beacon and
scheduled a process, it begins
monitoring its cache misses. Monitoring continues
until the loop-completion
beacon fires at the end of the loop. Then monitoring is reset
and paused until the next beacon event.

For each beacon process, the scheduler
regularly examines if the cache misses are less than the predicted
amount. When the misses exceed the
predicted upper bound CM(U) (see Equation~\ref{equation:LT7}),
Biscuit is alerted of a plausible attack on
the process, which is now treated as the victim.
% Because the scheduler
% has managed to schedule the processes such that cache contention 
% due to co-execution is minimal, the scheduler expects the 
% actual cache misses experienced by the process to be within the
% predicted upper bound CM(U), as in Equation~\ref{equation:LT7}.
To catch the plausible attacker and mitigate the attack, the 
scheduler conducts a combination of binary and linear search
on the executing processes,
as shown in Algorithm~\ref{Alg:mitigation}. 
% After scheduling all initial processes, the scheduler detects for 
% beacon calls. The beacon calls help the scheduler figure out the 
% necessary cache misses within the loop being executed at the moment. 
% At every beacon call, we run perf on the process to detect the 
% runtime cache misses for the loop. When we reach the beacon completion
% call for the loop, we stop the perf monitoring since the beacon complete 
% means the end of the loop. 
% At fixed intervals and at the beacon completion calls, the scheduler checks each 
% process's cache misses and compares it against the predicted cache misses 
% given by the beacon calls. When the scheduler detects the process's cache misses is 
% greater than the predicted cache misses, it assumes there is an attacker attacking
% the process. To figure out the attacker, the scheduler runs the mitigation
% algorithm.

The scheduler analyses the cache misses and isolates the attacker in the following manner.
The LLC is shared among the processes on the same
socket. The plausible attacker must therefore be executing on the same
socket as the victim. The scheduler searches for the plausible attacker
using misses per thousand instructions (MPKI) recorded over
a period of 10ms.
After recording the current MPKI, the scheduler de-schedules 
half of the executing processes (not including the victim)
% (which is now known due to its dynamic cache
% misses being higher than the predicted value).
For simplicity, 
we have two sets: the de-scheduled processes set (\textbf{set A}) and 
the scheduled processes set (\textbf{set B}). The 
scheduler checks if the new MPKI of the victim is less than
the previously recorded MPKI. If so, then the plausible attacker
must be in the de-scheduled process set (\textbf{set A}).
In such a scenario, the scheduler re-schedules \textbf{set A}
and continues with the victim process using this set.
If the plausible attacker were in the scheduled 
set (\textbf{set B}), then it continues with \textbf{set B} and
again carries out a split to isolate the attacker and victim
to the same subset. To avoid latency 
for de-scheduled processes, we always re-schedule the other
set immediately after the check; we know the set
does not have the plausible attacker, and cache contention is 
already resolved.
In short, after each iteration, the scheduler
reduces the number of suspect processes by half. The scheduler repeats
the above step until the number of suspect processes is less than
or equal to 8. Then it moves to a linear search.

Once we have 8 or fewer suspect processes, the scheduler conducts a 
linear search by recording MPKI over a period of 2ms instead of 10ms,
because only one process is de-scheduled at a time
between two readings of the MPKI. When the new 
MPKI is less than the previously recorded MPKI,
the de-scheduled 
process is the attacker. Otherwise, the scheduler re-schedules the
process back again and continues until all suspect processes are
checked. If the attacker is not isolated to any of these processes,
then Biscuit treats the victim itself as the plausible attacker.
That is, the attacker
is a beacon-enabled process masquerading as a victim. The 
victim process in this case fired a beacon with the predicted 
cache misses, but it
had been compromised, e.g. through a control data or non-control
data attack, which was leveraged to carry out the CSA. The 
scheduler is able to catch such attackers, as well. These cases
are checked in the Algorithm~\ref{Alg:mitigation} which describes the necessary details.

When all the
scheduled processes finish executing, the plausible attacker is
re-scheduled, because the process may have been wrongly classified as
an attacker (which might happen due to inaccuracies in cache miss
prediction). The process is allowed to progress only if no other
process is waiting to be scheduled.
If the
process does not complete, and other processes are in queue to 
be scheduled, then the plausible attacker will be de-scheduled again.
At this point, further forensic examination can be conducted on
the plausible attacker. Our results show that our attack detection
techniques are 100\% accurate on the cryptography benchmarks and
we have very few false positives for the DS attacks for San Diego
Vision benchmarks. The details of these are described in the evaluation.

% Since no other
% processes are scheduled, such a process should progress and complete. But 

%-------------------------------------------------------------------------------
\section{Evaluation}
\label{sec:evaluation}
%-------------------------------------------------------------------------------
Both side channel as well as degradation of service attacks were carried out in a multi-tenant environment. 
The experiments were conducted on a Dell PowerEdge R440 server,
which is equipped 
with an Intel Xeon Gold 5117 processor, clocked at 2.00 GHz.
The Dell PowerEdge R440 
has two sockets, each consisting of 14 cores and an
11-way associative 19 MB LLC.
We decided to carry out experiments with up to 18 jobs, utilizing
18 cores of the machine (a little over 50\% utilization, in accordance
with our threat model).
(Note that server farms typically have an utilization
of less than 50\% to effectively give SLA (service level agreement)
guarantees~\cite{overprovision}). Availability of some free cores ensured that the daemon processes were scheduled and did not interfere with the experimental setup. 
The server OS is Ubuntu 18.04 with 4.15 linux kernel.
Our baseline scheduler is the Completely Fair Scheduler (CFS),
which is the default linux scheduler. We use scikit-learn
to build the cache-miss model. We write the compiler passes in
LLVM 3.8 to collect the training data for the cache-miss model,
insert the model, and hoist the model.

\begin{figure*}
    \centering
    \includegraphics[width=0.90\textwidth]{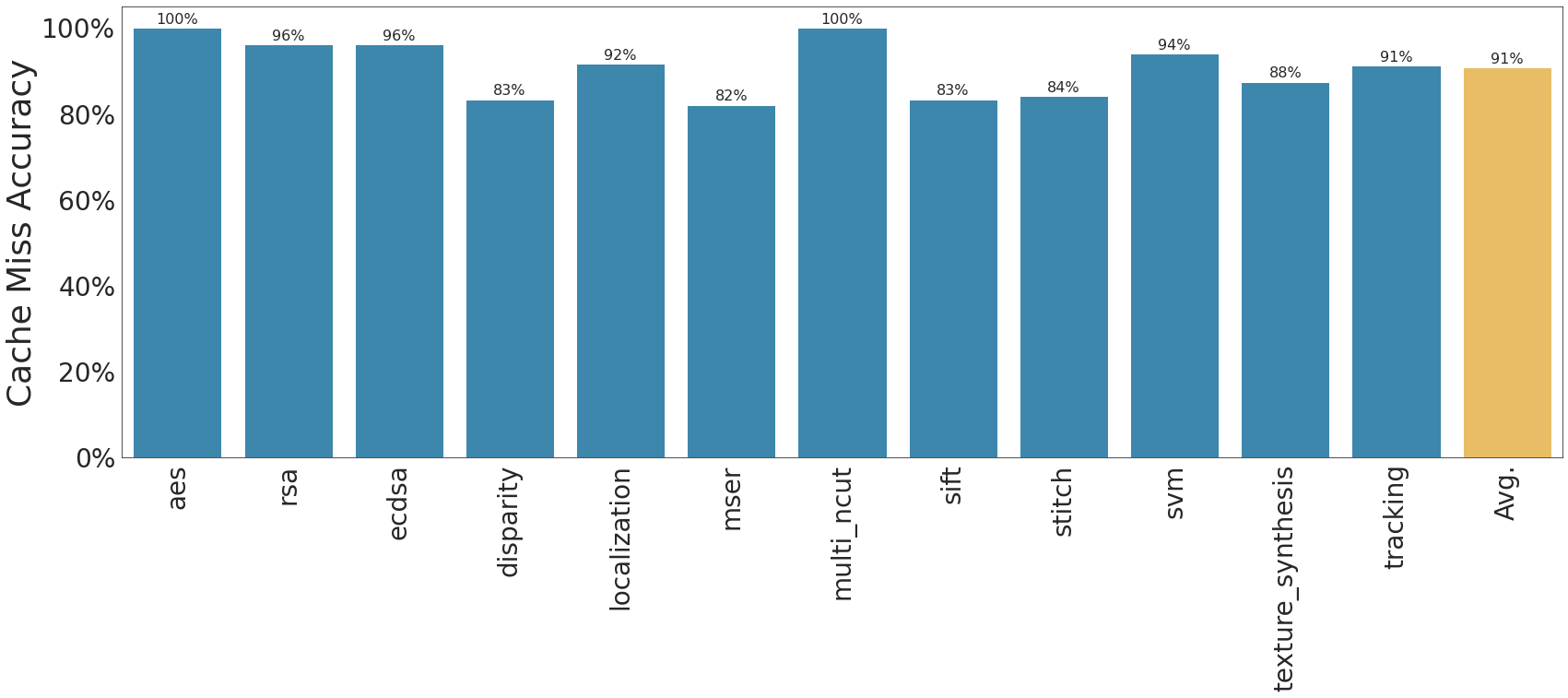}
    \vspace{-0.3in}
    \caption{Offline Cache Model Accuracy}
    \label{fig:accuracy}
\end{figure*}

\paragraph{Benchmarks:}
The cache-based side channel attacks (CSAs) that rely on cache attack techniques (CATs) have been successful
in extracting the secret keys of cryptography 
algorithms. We use OpenSSL's implementation of AES, RSA, 
and ECDSA to demonstrate the working of Biscuit in 
catching the CSAs. 
Each cryptography algorithm encrypts and then decrypts
random strings of random lengths with a random secret key.
To show that Biscuit can successfully detect and mitigate the 
CSAs in a multi-tenant 
environment, we run several image analysis and processing applications
from San Diego Vision Benchmarks Suite (SDVBS)~\cite{venkata2009sd}
alongside
the cryptography applications to simulate a real world job mix
that may be seen on server farms. 
We show that Biscuit catches CSAs on cryptography algorithms
with all three different CATs, namely Prime+Probe, Flush+Reload, and
Flush+Flush. In the case of non-crypto applications,
the CATs manage to degrade the
services of the victim by inducing a huge number of cache misses in
the victim. Biscuit can even detect such degradation of service
(DS) attacks. We also show the effectiveness of Biscuit against
degradation of service with Flush+Reload CAT on SDVBS.

% Typically applications for side channel attacks are extracting 
% cryptography algorithms' secret keys. To demonstrate biscuit's 
% defense and mitigation mechanisms, we used OpenSSL's implementations
% of AES, RSA, and ECDSA. These three cryptography algorithms have been shown to be 
% vulnerable to side channel attacks. For each cryptography algorithm, we 
% run numerous encryption of random strings with random lengths followed 
% by decryption while keeping the same secret keys. Aside from extraction 
% of data, the side channel attacks can lead to degradation of performance
% for programs. We augment these cryptography algorithms with vision 
% applications from San Diego Vision Benchmarks. The vision benchmarks 
% range from image analysis to image processing. Most of the applications
% are well-known so they could be a chance that the images can be extracted
% just like secret keys from cryptography algorithms based on the timing
% patterns. However, our focus is to see the degradation of these applications
% under the influence of side channel attacks. Together, both benchmarks 
% encompass a wide range of functionalities that side channel 
% attacks can be applied upon.

\paragraph{Attack Setup:}
We leveraged the CATs code from 
\href{https://github.com/IAIK/flush_flush}{https://github.com/IAIK/flush\_flush} 
that is based on Gruss et al.'s cache template attacks \cite{191010, flushflush}. 
The CATs can be used to monitor cache accesses of specific locations in shared libraries. 
By looking at the accesses, one can infer information such as secret keys in cryptographic algorithms \cite{yarom2014recovering, flushreload}.
CATs have three stages: monitoring, logging, and analysis.
(Thus, our attack does not have an inferring stage, but it can be 
done offline by looking at the accesses).
For our purposes, the monitoring step is the most important, because
it causes the cache misses.
% In other words, in a CAT attack scenario, an attacker is 
% continuously monitoring the victim until the victim is active.
% We thus focus on this stage with
% an aim to stop the attack.
For the shared libraries, we used the libcrypto library 
for OpenSSL, and we created a libsdcommon library from 
the common folder consisting of general image functions for SDVBS.
Each cryptographic algorithm has its own specific execution accesses
in the libcrypto library.
We profiled the libraries to find the necessary library code address 
range that is needed to figure out the secret key in each cryptographic algorithm. 
In the case of SDVBS, the attack should degrade the performance, so we just monitor common functions in the 
library used by all SDVBS benchmarks.

\paragraph{Evaluation Method:}
We create different configurations, consisting of different
applications, ranging from 3 to 18 processes to test the Biscuit
scheduler versus CFS. 
We first test the scheduler efficiency in
terms of the performance
of the applications scheduled without the attacker. Note that even
without the attacker, the Biscuit scheduler monitors the cache misses
and regularly checks for attacks. Then we test the ability of
the Biscuit scheduler to catch the attack. 
In our experiments, there is only one attacker and one victim.
The remaining processes are benign. This mimics a typical attack scenario.
In these experiments, we add a CSA to the configuration.
When demonstrating an attack over OpenSSL, we use applications
from SDVBS as co-executing applications and vice-versa. For example,
to test an attack on AES in a configuration of three processes, we
used SVM from SDVBS and the attack as the co-executing process.
%For recording 
%performance we did not include the attack and time the scheduler 
%from the start of the processes till the end of all processes. 
% When timing the configuration with attack, we stop the timer when all 
% processes except the attack completes because the attacker does not
% complete soon comparatively.
%Below, we also compare the total cache misses with and without the attack.
%Biscuit scheduler is expected to catch the attack and hence must have
%lower cache-misses than CFS in configurations with an attack.

% We create different configurations that we run using both the biscuit scheduler 
% and CFS. In each configuration, we have only one victim while the rest of
% the processes will be of other applications. For OpenSSL algorithms as victims, we
% use SDVBS as the other processes and vice versa. We take this configuration and 
% run it twice: one with the side channel attack and once without the side channel attack. 
% For example, in AES with three processes, we will have one configuration containing an 
% AES process and a SVM process from SDVBS and another configuration containing the 
% same processes alongside an attacker like Flush and Reload. We time the scheduler 
% from once it starts until all processes are completed except the attacker. The reason 
% being the attacker takes way longer compared to all the processes so it will always 
% be the last process and will decide the final time. We just want to compare the cache misses 
% it takes for all processes to finish in the presence of an attacker or not.

\subsection{Cache Model Accuracy}

\begin{table}[]
 \centering
 \begin{tabular}{|l|l|l|}
   \hline
   \textbf{Benchmarks} & \textbf{F-Score} & \textbf{False Positive Rate}\\
    \hline
   OpenSSL  &	 1 & 0\%  \\
   SDVBS    &    0.9375 & 5\%   \\
    \hline
  \end{tabular}
  \caption{F-Score}
  \label{table:fpfnssl}
\end{table}

The cache-model accuracy depends on the training data and the types
of loops. For OpenSSL, we use random strings of different lengths and
random secret keys for both training and testing.
For SDVBS, we leverage the input data sets provided with the
suite for training and testing. The model predicted
value is further compromised in the case of expected beacons in which
the loop bounds used in the model are the expected loop bound obtained
during training, and not the actual iterations. Because we use the
standard deviation ratio to predict the cache misses (see
Equation~\ref{equation:LT7}), we account for the
$k*CM$ to get the error in prediction with test data.
The average accuracy is 91\%, as shown in Figure~\ref{fig:accuracy}.

% The actual cache misses can be above or below these predicted
% cache misses but we take this into account by bringing in the standard deviation constant.
% However, there are other issues in the data that cause the model not to be exact. As a result,
% the accuracy shown in Figure~\ref{fig:accuracy} has an average 91\%.
% We got these accuracies
% by running our script by dividing the data into training and test data. 

In each OpenSSL application, random strings of various lengths
are encrypted multiple times.
The number of times the algorithm runs
is different in the case of training versus testing.
For example, one application
was run 5,000 times during training versus 7,500 times
during testing. While the accuracy is 100\% for AES, it is 96\% for
RSA and ECDSA.

During the runs with the Biscuit scheduler,
all attacks on OpenSSL were caught. The attacker caused
high cache misses in the victim
even when the cache was
just being warmed up for the loop. Once the 
attacker was caught, the cache was already warmed up,
and the execution continued normally.
% Therefore all CSAs are caught by
% the Biscuit scheduler, and the attacker is isolated using the
% scheduling Algorithm~\ref{Alg:mitigation}
% described earlier, without
% besmirching a normal process. In other words, all CSAs are caught
% with no false positives and false negatives, leading to an F-Score
% of 1.
There were no false positives or negatives, giving an F-Score of 1.

% For OpenSSL algorithms, we run each encryption followed by decryption many times. In each run, 
% we set the number of times that the encryption and decryption algorithms run. In following runs,
% we increase the number of times that the algorithm runs. These different runs encompasses our
% training set. For the test set, we do the same thing but we use different values for the number
% of times that the algorithm runs. For example, one train run had 5000 encryptions and one test
% run had 7500 encryptions. The accuracies for these algorithms are above 91\% but the model
% is not 100\% for RSA and ECDSA. However, at runtime, the attacker is always caught 
% because the attacker is increasing the cache misses while the cache is warming 
% up for the loop leading to a way higher number of cache misses. Once the attacker is caught, 
% the cache has warmed up so the misses will be within the normal amount that is less than the
% predicted cache misses. From this point, all processes execute normally
% without any de-schedules. Therefore, OpenSSL has no false positives and false negatives meaning an 
% F-Score of 1. 

SDVBS has different image sizes: test, sim, sqcif, qcif, cif, vga, 
and fullhd. We mostly use cif and vga for training and  test on fullhd
because these sizes are large enough to cause at least some cache
misses for each loop.
(In contrast, test, sim, sqcif, and qcif have such small loop bounds
that they execute without causing much cache misses.)
% almost always zero cache misses due to small loop bounds. However, some loops in cif and vga
% have 0 cache misses and they are 1000+ misses in fullhd. 
However, three applications (localization, svm, and multi\_ncut) do
not have fullhd or vga data. Hence, localization is
trained on qcif and cif and tested on vga. Svm and multi\_ncut
are trained on sqcif and qcif and tested on cif. Because image sizes
are small, svm finishes too fast for the attacker to attack the
process, whereas, multi\_ncut takes several minutes for the
cif data image, so this benchmark is easily attacked.
Due to these issues (especially the small input sizes for SDVBS),
the training for each benchmark leads to fluctuations in the
accuracy of the cache miss models. The accuracy ranges from 83 to 90\%.
Due to relatively lower prediction accuracy, false 
positives occurred when co-executing alongside more than 12 processes,
giving an F-score of 0.9375. There are six configurations that tend to produce 
false positives. These configurations sometimes de-schedule a non-attack process when there is no attacker.
Such a process can, however, continue execution on a quarantined socket (somewhat compromising consolidation).
It may be noted the, both the F-scores (1 for RSA and 0.9375 for SDVBs) and false positive rate (0\% and 5\% for RSA and SDVBs respectively) for our approach are better than the worst reported values in SpyDetector (F-score of 0.83 and false positive rate of 30\% ) and Cloudradar (False positive rate of 20-30\%). 

\subsection{Detection Efficiency}

The schedulers' efficiency in catching these side channel attacks
is determined by {\it how early} the attacker is caught; the earlier 
an attack is caught, the better, since this limits
the attacker's ability to extract more information on the secret key.
In other words, the scheduler must thwart an attack before the
secret data is retrieved by the attacker. The efficiency in 
thwarting the attack can be described through the amount of data that was 
was allowed to leak before the attack is caught and mitigated.
With this motivation, we define the detection efficiency as the
complement of the fraction of the data leaked with Biscuit versus
without Biscuit. The data recorded by the CATs corresponds to the cache lines
that were accessed by the victim and the cycles elapsed between the accesses.
Since cache access pattern is recorded during the entire execution
of the victim without Biscuit, the detection efficiency can be
defined as the complement of the fraction of time the attacker 
executed with Biscuit versus without Biscuit, i.e.,
\begin{equation}\label{eq:detection}
   D = 1 - \frac{\texttt{Attack active time when thwarted}}
   {\texttt{Attack active time when successful}}
\end{equation}
where D is the Detection Efficiency of the defense mechanism
and Attack active time is the duration from when the attack started to
the time when it either completed or was de-scheduled.
% Since all processes are started simultaneously, the above equation
% signifies the ratio of time the victim executed relieved of the attacker, or in other words the time
% the victim was not under attack after the adversary started executing.
According to this equation and our assumption that the attack was
successful with the baseline Linux scheduler CFS, the detection efficiency of CFS is 0\%. Thus, the higher the detection efficiency is, the better, because in such cases, the attack is not allowed to progress much, and the scheduler catches the attack process early on, minimizing the fraction of active time when attacker partially stole the secret key. 

\begin{figure}
    \centering
    \includegraphics[width=0.46\textwidth]{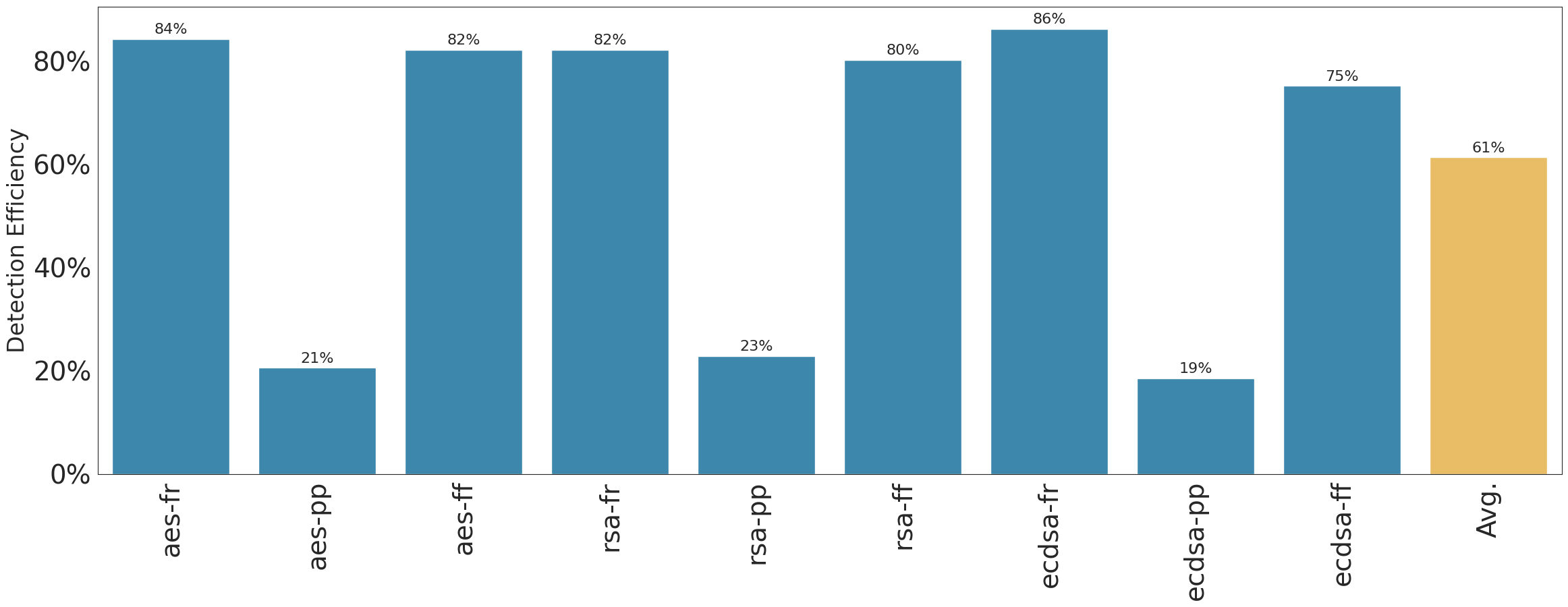}
     \caption{Detection Efficiency for OpenSSL}
    \label{fig:detection1}
\end{figure}

In the case of CSAs on OpenSSL, as shown
in Figure~\ref{fig:detection1}, we can see that the attack was
caught early in all the algorithms except Prime+Probe. This is because
Prime+Probe causes cache misses in the victim very slowly compared to
Flush+Reload or Flush+Flush. Also, comparatively Prime+Probe
utilizes few memory addresses, because more memory addresses
mean more eviction sets that take memory space and time to initialize.
On average, the scheduler detects the CSA
with 61\% detection efficiency, meaning that no more than 39\% of the full attack was allowed to proceed.

% One thing to notice, the OpenSSL algorithms are enclosed within
% a loop nest so there is only one cache model to encompass
% the entire loop. This loop nest determines both the runtime and cache misses
% so the attacks lead to this one loop increasing its misses at a 
% fast pace.

\begin{figure}
    \centering
    \includegraphics[width=0.46\textwidth]{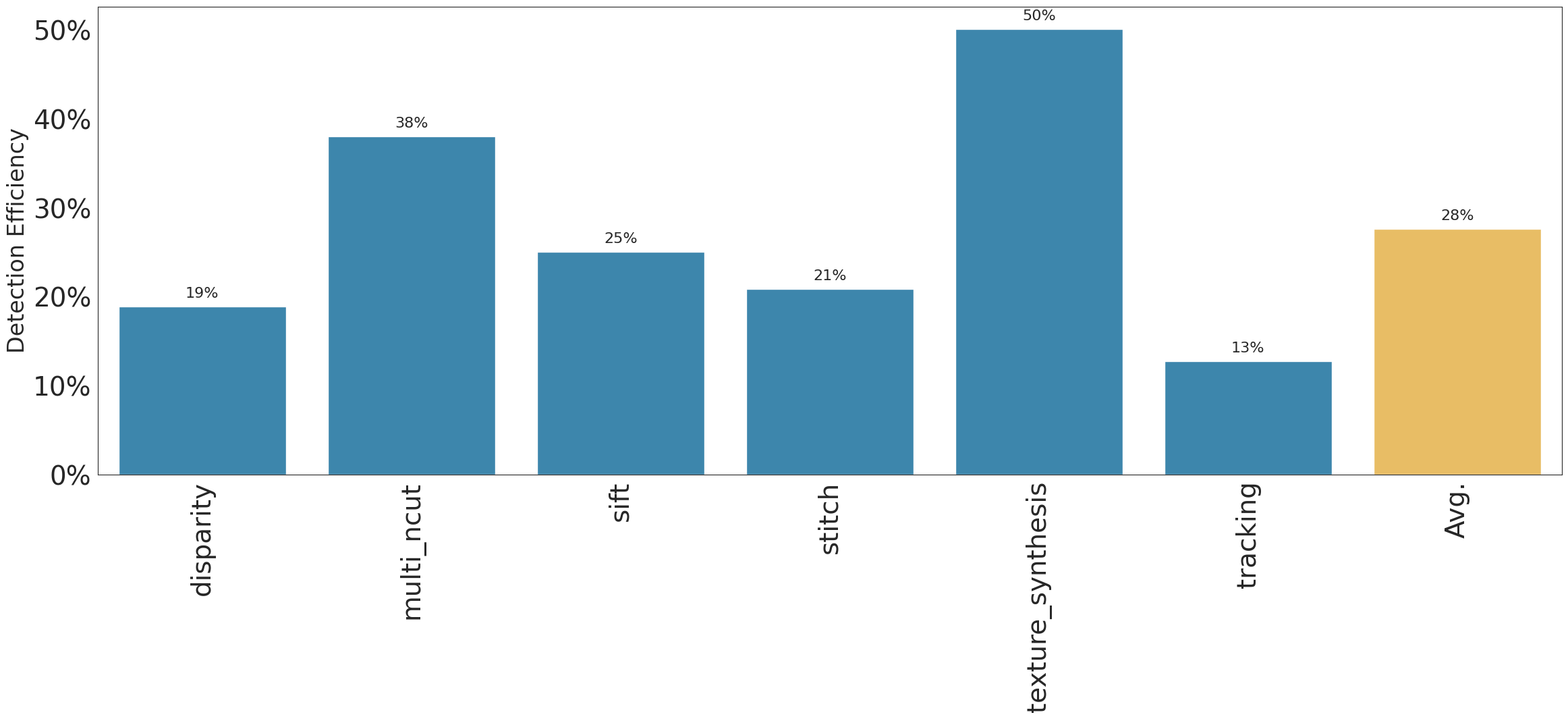}
    \caption{Detection Efficiency for SDVBS}
    \label{fig:detection2}
\end{figure}

In the case of a DS attack, Equation~\ref{eq:detection} is
also applicable, because the goal
is to keep the degradation of service to a minimum.
In the case of a DS attack on SDVBS (see Figure
~\ref{fig:detection2}), the detection efficiency is low due to
two reasons. The first reason is the shared library's attack memory
lines may not be used frequently by the victim. The beacon is hoisted
at the outer loop; the misses take time to
accumulate; and so the attack gets detected late.
The second reason is the program has a short execution time, so
the attack is detected very late because it needs to accumulate
misses.
For the few cases in which the cache model is less accurate,
the attack can still be detected because the predicted misses are
exceeded, and the scheduler starts mitigation using MPKI.
However, in a few cases the program executes
too fast for any detection and mitigation of the attacker.
Three benchmarks in particular,
mser, svm, and localization, finished executing even before
the attacker caused cache-misses. On average, the detection efficiency
is 28\%, but the scheduler is still able to isolate the culprit.

% The best benchmark is texture synthesis

\begin{figure*}
\begin{subfigure}{0.32\textwidth}
\centering\includegraphics[width=\textwidth]{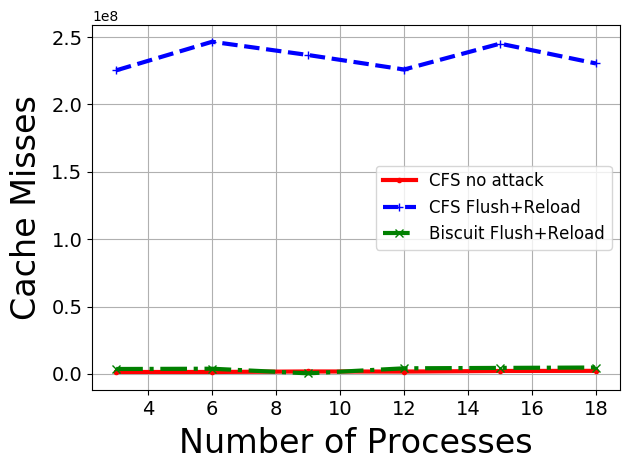}
 \caption{Flush+Reload Misses}%\label{fig:pLoop4}
\end{subfigure}\hfill
\begin{subfigure}{0.32\textwidth}
\centering\includegraphics[width=\textwidth]{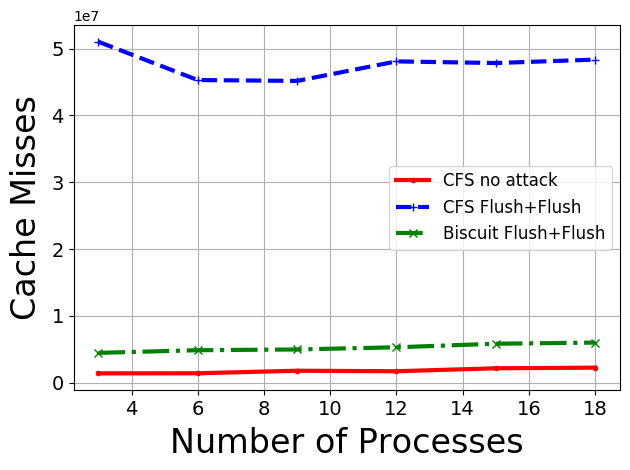}
 \caption{Flush+Flush Misses}%\label{fig:pLoop4}
\end{subfigure}\hfill
\begin{subfigure}{0.32\textwidth}
\centering\includegraphics[width=\textwidth]{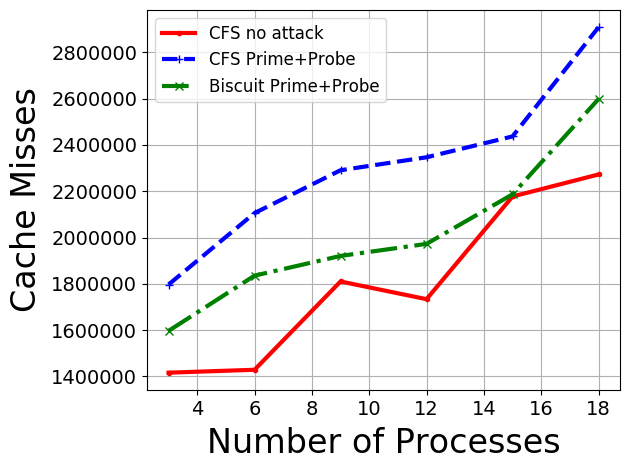}
 \caption{Prime+Probe Misses}%\label{fig:pLoop4}
\end{subfigure}\hfill
\caption{RSA Misses}%\label{fig:pLoop4}
\label{fig:RSAMISSES}
\end{figure*}

\begin{figure*}
\centering
\begin{subfigure}{0.32\textwidth}
\centering\includegraphics[width=\textwidth]{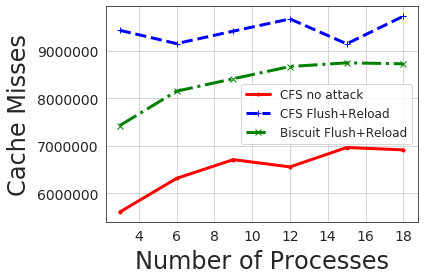}
 \caption{Stitch Misses}%\label{fig:pLoop4}
\label{fig:StitchMisses}
\end{subfigure}\hfill
\begin{subfigure}{0.32\textwidth}
\centering\includegraphics[width=\textwidth]{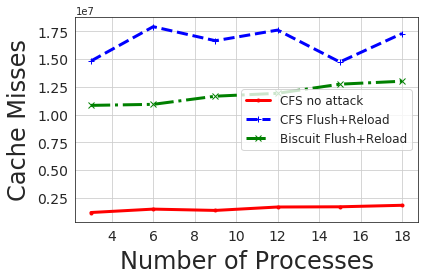}
 \caption{Texture Synthesis Misses}%\label{fig:pLoop4}
\label{fig:TextureMisses}
\end{subfigure}\hfill
\begin{subfigure}{0.32\textwidth}
\centering\includegraphics[width=\textwidth]{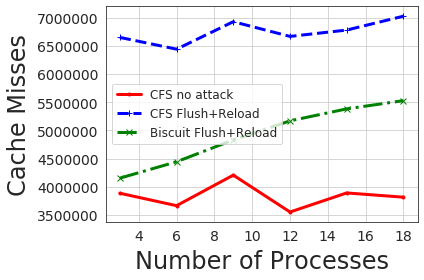}
 \caption{Multi-NCut Misses}%\label{fig:pLoop4}
\label{fig:MCMisses}
\end{subfigure}\hfill
\caption{SDVBS Cache Misses}
\label{fig:SDVBSMISSES}
\end{figure*}

\begin{figure*}
\centering
\begin{subfigure}{0.32\textwidth}
\centering\includegraphics[width=\textwidth]{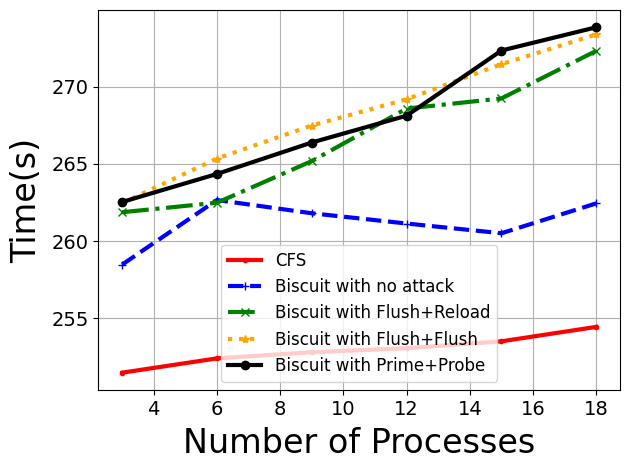}
 \caption{AES Time}%\label{fig:pLoop4}
\label{fig:AESTime}
\end{subfigure}\hfill
\begin{subfigure}{0.32\textwidth}
\centering\includegraphics[width=\textwidth]{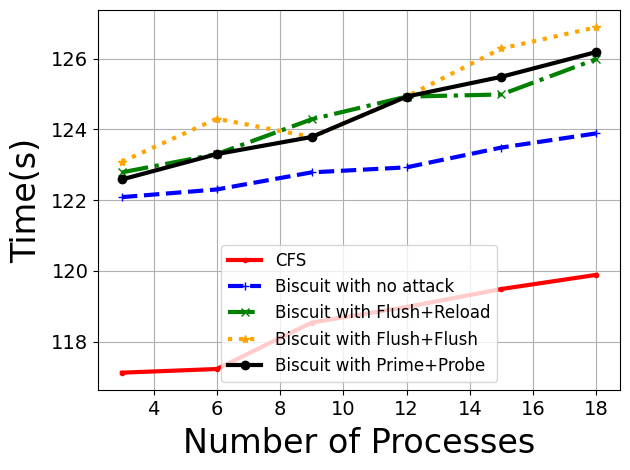}
 \caption{RSA Time}%\label{fig:pLoop4}
\label{fig:RSATime}
\end{subfigure}\hfill
\begin{subfigure}{0.32\textwidth}
\centering\includegraphics[width=\textwidth]{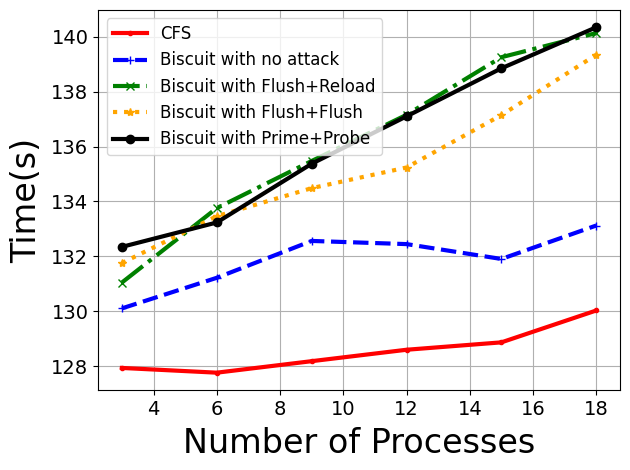}
 \caption{ECSDA Time}%\label{fig:pLoop4}
\label{fig:ECSDATime}
\end{subfigure}\hfill
\caption{Timing between CFS and Biscuit for OpenSSL Algorithms alongside no attacks }
\label{fig:SSLTiming}
\end{figure*}

\begin{figure*}
\centering
\begin{subfigure}{0.32\textwidth}
\centering\includegraphics[width=\textwidth]{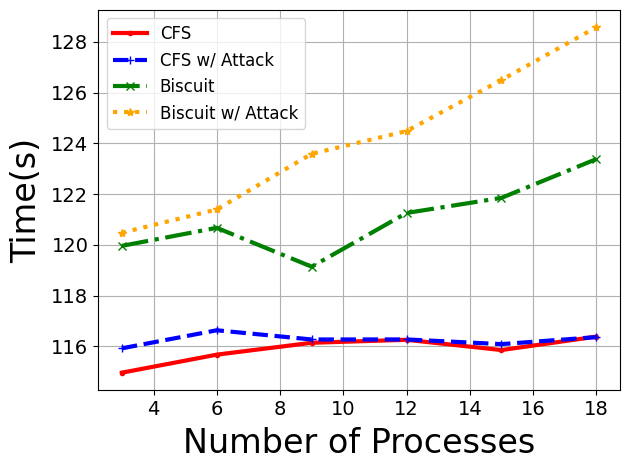}
 \caption{Stitch Time}%\label{fig:pLoop4}
\label{fig:StitchTime}
\end{subfigure}\hfill
\begin{subfigure}{0.32\textwidth}
\centering\includegraphics[width=\textwidth]{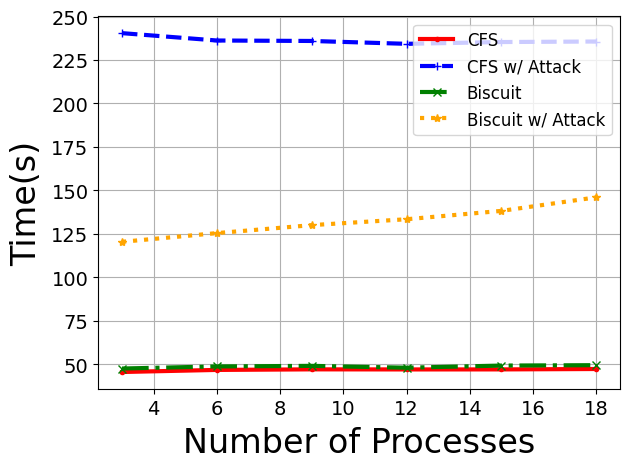}
 \caption{Texture Synthesis Time}%\label{fig:pLoop4}
\label{fig:TextureTime}
\end{subfigure}\hfill
\begin{subfigure}{0.32\textwidth}
\centering\includegraphics[width=\textwidth]{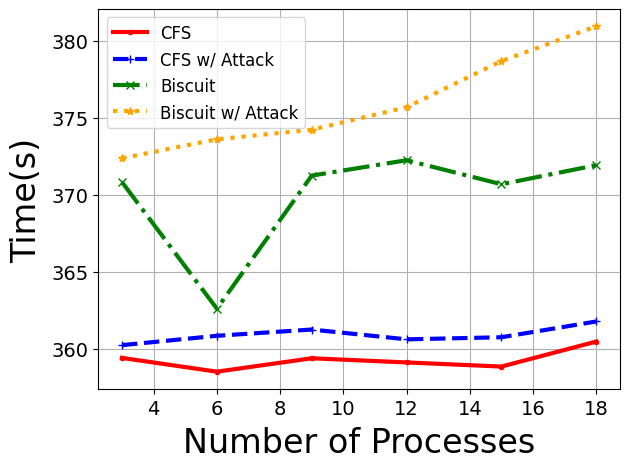}
 \caption{Multi-NCut Time}%\label{fig:pLoop4}
\label{fig:MCTime}
\end{subfigure}\hfill

\caption{Timing between CFS and Biscuit for SDVBS Algorithms alongside no attacks }
\label{fig:SDVBSTiming}
\end{figure*}

\subsection{Performance}
\begin{figure*}
\begin{subfigure}{0.32\textwidth}
\centering\includegraphics[width=\textwidth]{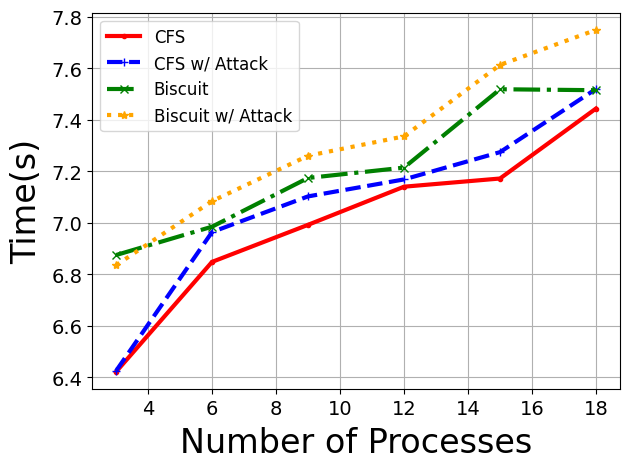}
 \caption{Localization Time}%\label{fig:pLoop4}
\label{fig:LocalTime}
\end{subfigure}\hfill
\begin{subfigure}{0.32\textwidth}
\centering\includegraphics[width=\textwidth]{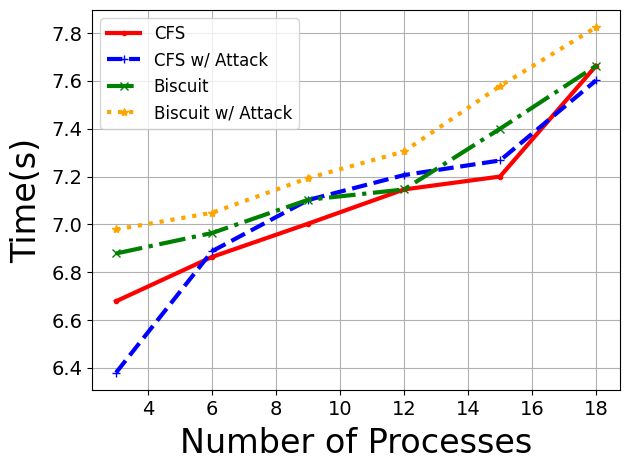}
 \caption{Mser Time}%\label{fig:pLoop4}
\label{fig:MserTime}
\end{subfigure}\hfill
\begin{subfigure}{0.32\textwidth}
\centering\includegraphics[width=\textwidth]{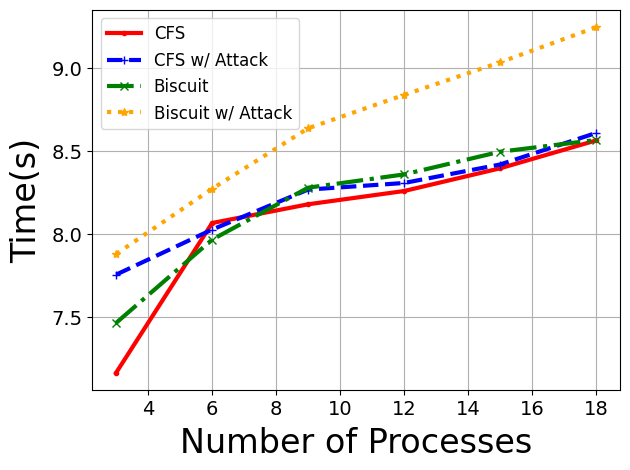}
 \caption{Sift Time}%\label{fig:pLoop4}
\label{fig:SiftTime}
\end{subfigure}\hfill

\begin{subfigure}{0.32\textwidth}
\centering\includegraphics[width=\textwidth]{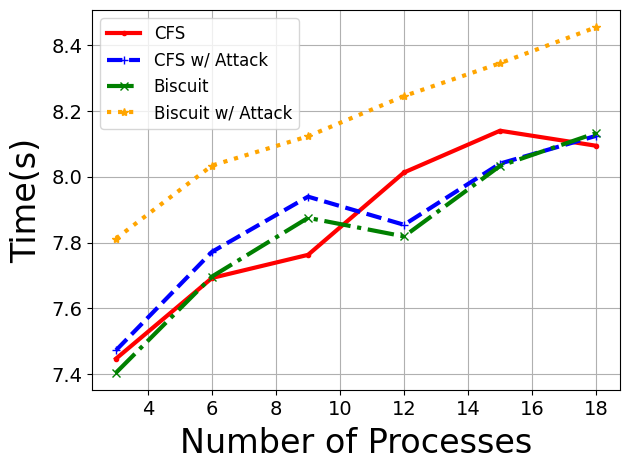}
 \caption{Disparity Time}%\label{fig:pLoop4}
\label{fig:DisTime}
\end{subfigure}\hfill
\begin{subfigure}{0.32\textwidth}
\centering\includegraphics[width=\textwidth]{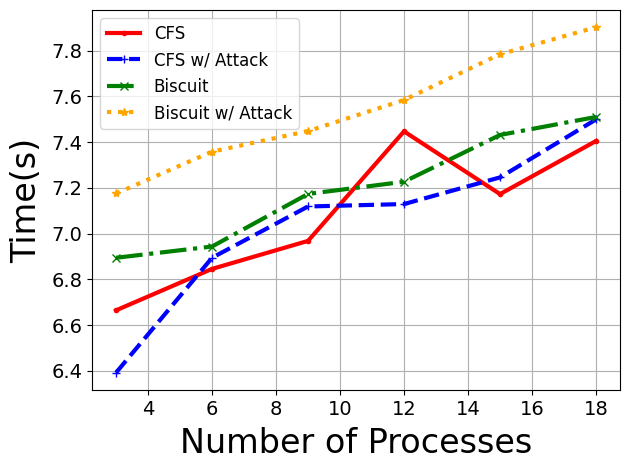}
 \caption{SVM Time}%\label{fig:pLoop4}
\label{fig:SVMTime}
\end{subfigure}\hfill
\begin{subfigure}{0.32\textwidth}
\centering\includegraphics[width=\textwidth]{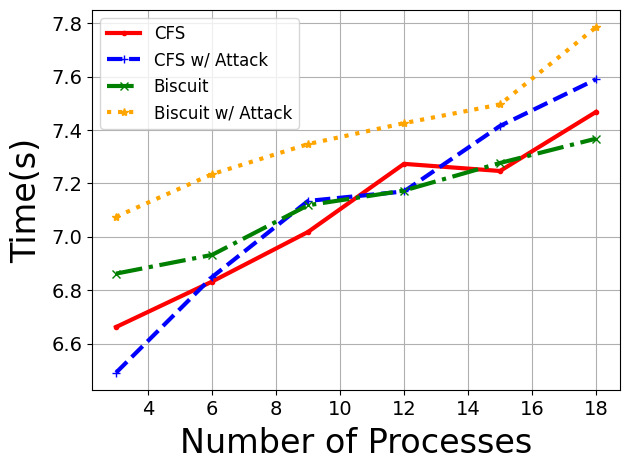}
 \caption{Tracking Time}%\label{fig:pLoop4}
\label{fig:TrackingTime}
\end{subfigure}\hfill

\caption{Timing between CFS and Biscuit for SDVBS Algorithms alongside no attacks }
\label{fig:SDVBSTiming2}
\end{figure*}

The timing differences between Biscuit and CFS
for OpenSSL and several SDVBS benchmarks are shown in 
Figure~\ref{fig:SSLTiming}, Figure~\ref{fig:SDVBSTiming}, and Figure~\ref{fig:SDVBSTiming2}. The 
Biscuit scheduler is almost as efficient as CFS.
This can be easily seen in short programs
such as SVM, which shows
the two curves being interchangeable. Without any attacker, 
CFS is better by a maximum of 6\%.
This 6\% overhead is due to Biscuit's constant monitoring of cache
misses and logging even when there are no attacks. When an attacker is present in a CFS scheduling setting, most victims incur 5-10\% overhead, however they incur 1.5-2.5x cache misses. In the case of the benchmark ``texture synthesis'' with
CFS and an attack, the increase in time and cache misses are 5x and
7x, respectively (see Figures~\ref{fig:TextureTime} and \ref{fig:TextureMisses}). However, with Biscuit the overhead and
cache misses are reduced to 3x and 4x. The huge increase in overhead is due to the attacker, and Biscuit causes a 40\% decrease in the overhead by de-scheduling the attacker. For the other benchmarks running alongside Biscuit and an attacker, the victim experiences an average overhead of 8.1\%.
In the presence of attackers, Biscuit slows down processes 
during the mitigation phase by de-scheduling and re-scheduling 
the processes to find the plausible culprit causing the overhead.
The worst-case overhead, excluding the texture synthesis benchmark,
is 11\%, due to the misclassification of a 
process as an attacker, which caused it to be re-scheduled later. 
% the overhead is at worse 11\% due to the de-scheduling and
% rescheduling of processes during the mitigation phase and processes 
% being mis-classified as attacker when they are not and getting 
% de-scheduled.
% In terms of degradation, most processes increase time but it is small
% around 5-10\% increase in time compared to the 2-5x increase in misses
% except for the texture synthesis benchmark in SDVBS. As seen in
% Figure~\ref{fig:TextureTime}, it increased by 4x in time compared to
% the increase of 7x see in Figure~\ref{fig:TextureMisses}. When we add
% Biscuit, it figured out the attacker and de-scheduled it causing the
% texture synthesis process to be faster. 

\begin{figure*}
\begin{subfigure}{0.32\textwidth}
\centering\includegraphics[width=\textwidth]{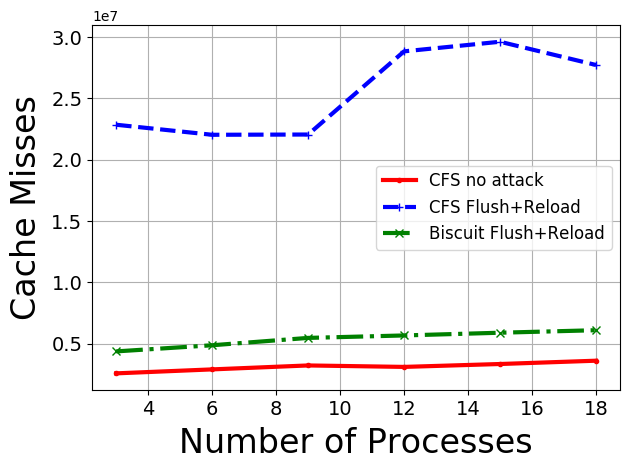}
 \caption{Flush+Reload Misses}%\label{fig:pLoop4}
\end{subfigure}\hfill
\begin{subfigure}{0.32\textwidth}
\centering\includegraphics[width=\textwidth]{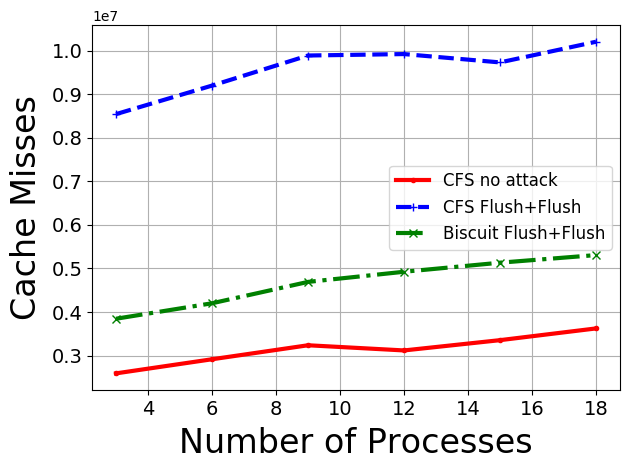}
 \caption{Flush+Flush Misses}%\label{fig:pLoop4}
\end{subfigure}\hfill
\begin{subfigure}{0.32\textwidth}
\centering\includegraphics[width=\textwidth]{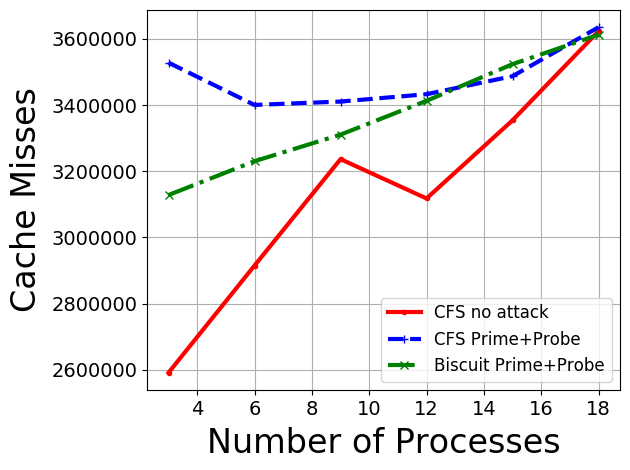}
 \caption{Prime+Probe Misses}%\label{fig:pLoop4}
\end{subfigure}\hfill
\caption{AES Misses}%\label{fig:pLoop4}
\label{fig:AESMiss}
\end{figure*}

\begin{figure*}
\begin{subfigure}{0.32\textwidth}
\centering\includegraphics[width=\textwidth]{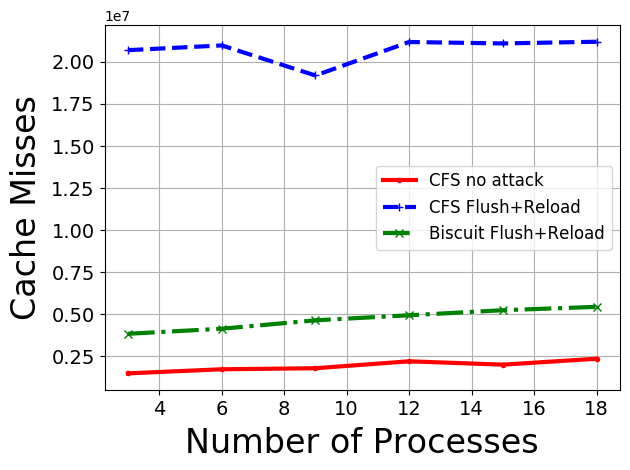}
 \caption{Flush+Reload Misses}%\label{fig:pLoop4}
\end{subfigure}\hfill
\begin{subfigure}{0.32\textwidth}
\centering\includegraphics[width=\textwidth]{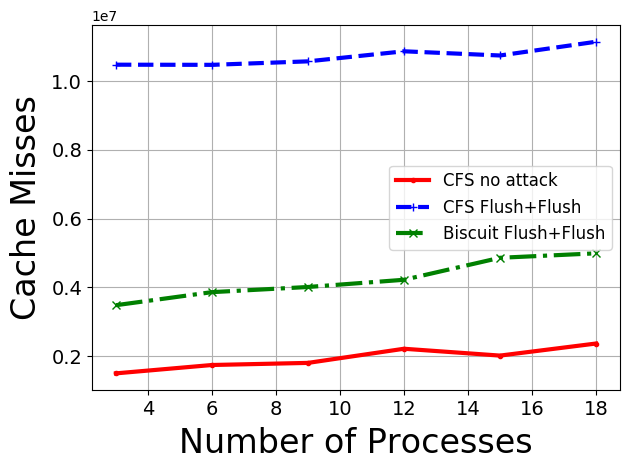}
 \caption{Flush+Flush Misses}%\label{fig:pLoop4}
\end{subfigure}\hfill
\begin{subfigure}{0.32\textwidth}
\centering\includegraphics[width=\textwidth]{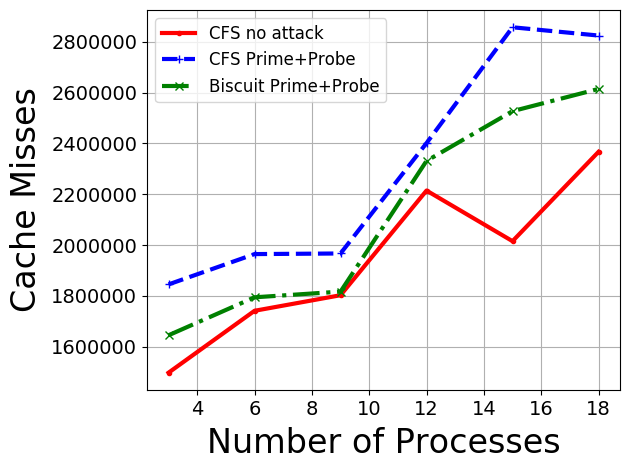}
 \caption{Prime+Probe Misses}%\label{fig:pLoop4}
\end{subfigure}\hfill
\caption{ECDSA Misses}%\label{fig:pLoop4}
\label{fig:ECDSAMISSES}
\end{figure*}

\begin{figure*}
\centering
\begin{subfigure}{0.32\textwidth}
\centering\includegraphics[width=\textwidth]{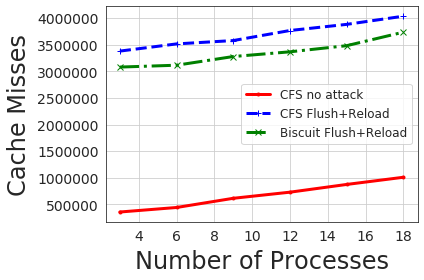}
 \caption{Localization Misses}%\label{fig:pLoop4}
\label{fig:LocalMisses}
\end{subfigure}\hfill
\begin{subfigure}{0.32\textwidth}
\centering\includegraphics[width=\textwidth]{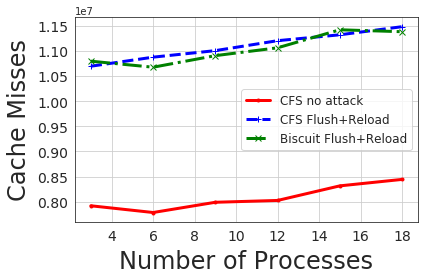}
 \caption{Mser Misses}%\label{fig:pLoop4}
\label{fig:MserMisses}
\end{subfigure}\hfill
\begin{subfigure}{0.32\textwidth}
\centering\includegraphics[width=\textwidth]{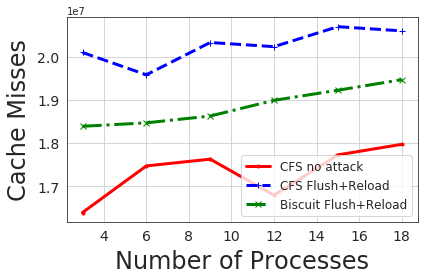}
 \caption{Sift Misses}%\label{fig:pLoop4}
\label{fig:SiftMisses}
\end{subfigure}\hfill

\begin{subfigure}{0.32\textwidth}
\centering\includegraphics[width=\textwidth]{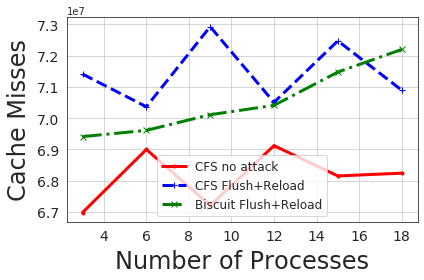}
 \caption{Disparity Misses}%\label{fig:pLoop4}
\label{fig:DisMisses}
\end{subfigure}\hfill
\begin{subfigure}{0.32\textwidth}
\centering\includegraphics[width=\textwidth]{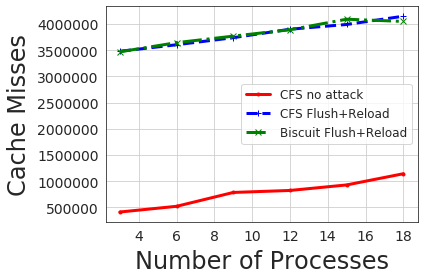}
 \caption{SVM Misses}%\label{fig:pLoop4}
\label{fig:SVMMisses}
\end{subfigure}\hfill
\begin{subfigure}{0.32\textwidth}
\centering\includegraphics[width=\textwidth]{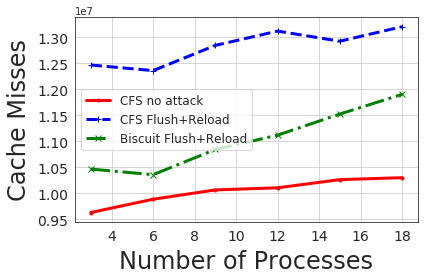}
 \caption{Tracking Misses}%\label{fig:pLoop4}
\label{fig:TrackingMisses}
\end{subfigure}\hfill

\caption{Rest of SDVBS Cache Misses}
\label{fig:SDVBSMISSES2}
\end{figure*}

% However, Biscuit is more efficient than the CFS when it comes to monitoring under attack. 
As mentioned before, Biscuit catches the cache
attacks in all the benchmarks. Since mser, svm, and localization
execute for a very short time, these benchmarks
could not be attacked. The difference in cache misses between
the Biscuit scheduler (which is actively detecting and mitigating
the attacks) and CFS (which does not do any detection nor mitigation) is
shown in Figure~\ref{fig:RSAMISSES}, Figure~\ref{fig:SDVBSMISSES}, Figure~\ref{fig:AESMiss}, Figure~\ref{fig:ECDSAMISSES}, and Figure~\ref{fig:SDVBSMISSES2}. 
In the case of OpenSSL, the cache misses decrease by 5x,
except for the Prime+Probe attack, which had its shortcomings
as explained before. In SDVBS, the longer
programs also had a larger decrease in their cache misses.
For mser, svm, and localization, the cache misses are identical to
CFS because there is no attack due to them completing very fast.
Other observations include the lines in the figures overlapping, and
the Biscuit scheduler sometimes having more
misses than CFS with an attack in SDVBS. The overlapping lines,
like in Figure~\ref{fig:RSAMISSES} for Flush+Reload, is
not actually an overlap. The two graphs have different values, but
the CFS with attack has a lot of misses, causing the other two lines to
get combined based on the y-axis. The Biscuit scheduler has more misses 
in the few cases in which a process got falsely flagged as an attack
and de-scheduled. 
Once a process is re-scheduled, it has to rewarm its cache, creating
more misses. The other reason processes with Biscuit can have more
misses is the non-determinism of a program's cache misses,
like in Prime+Probe for AES.

In summary, Biscuit 
defends against CSA and DS attacks with an overhead of less
than 6\% during normal scenarios (no attack) and 11\% during attack scenarios which is an acceptable overhead.
%-------------------------------------------------------------------------------
\section{Related Work}
\label{sec:relatedwork}
%-------------------------------------------------------------------------------
Refer to the introduction for the most closely related work.
%Here we
%broadly categorize other works based on software or hardware approaches, and
%further divide them based on the methods (such as prevention, detection, or
%mitigation).
Here we provide a broad overview of other techniques.

``Cache-mapping'' partitions the cache to make it harder for attackers to figure
out addresses which belong to the same set.
Methods such as~\cite{qureshi2018ceaser, encryptedcache} encrypt the memory
addresses using secret keys that are stored in isolated memory. This requires
additional memory, and the key must be changed regularly. 
RPCache~\cite{wang2007new}
and NewCache~\cite{7723806} use randomness in the cache replacement policy to
prevent the attacker from mapping the cache sets, but they do not scale with the last-level cache. 

Numerous techniques are based around cache lines within a set.
PLcache \cite{wang2007new} employs ``line-locking'', which locks cache lines
so that attackers cannot evict other processes. Similarly, Nomo~\cite{nomo}
restricts processes to only a few lines within a cache set. 
In contrast to these hardware solutions, DAWG~\cite{8574600} and CATalyst~\cite{catalyst} use 
software mechanisms to do way-partitioning.
Note that all these techniques limit the number 
of cache lines for each process within the same set, which can
lead to degradation.
%leading to some degradation 
%when processes need to use the same cache set.
Intel Cache Allocation Technology~\cite{herdrich2016cache}, used 
as a mechanism for way-partitioning, does not duplicate shared libraries for all partitions, so side channel
attacks can still detect hits of accesses to shared libraries. Since the flush instruction allows a process 
to remove shared memory accesses, Flush+Flush and Flush+Reload can still be used to attack the victim.
In a similar fashion, SHARP~\cite{sharp} only evicts cache lines that are not available
in the L1 or L2 caches of any process. This method requires changing
hardware to allow more communication among L1, L2, and LLC.
In addition, SHARP throws an alarm which
needs to be caught by the OS and can be a false positive.

% Timing solutions
Some methods require disrupting the time signals so that the attacker cannot
use the timing methods. One way is to take programs and 
disrupt the cycles by buffering extra instructions~\cite{wu2018eliminating}. 
The two programs are identical in terms of functionality, but the extra 
instructions means slower execution. Other methods add noise frequently to
programs~\cite{zhang2013duppel}. This method works for L1 caches but not for
LLCs, because the attacker is basing his assumption on the virtual addresses. Just like cache-mapping methods, some methods like TimeWrap~\cite{6237011} and TSCache~\cite{10.1145/3195970.3196003} add some randomization to timing signals, making it harder for attackers to infer the cache's state.

These mechanisms, like those compared against in the
introduction, either require hardware changes, or implement
hard isolation, which leads to cache reservation and many false
positives and negatives. Biscuit
solves these problems with a cache-miss model tailored at the granularity 
of loops, and with a scheduler that first schedules efficiently and then detects
and mitigates any cache-based side channel attacks and degradation of service. In particular, since Biscuit is a compiler-based solution, it can precisely compute the cache footprints (and corresponding expected cache misses) using the runtime values of loop bounds, which in turn leads to zero false positives.

%-------------------------------------------------------------------------------
\section{Conclusion}
\label{sec:conclusion}
%-------------------------------------------------------------------------------

In this work, we demonstrate a compiler-guided scheduler called Biscuit,
which detects the cache-based side channel attacks for processes on
multi-tenancy server farms with very high accuracy.
We show that Biscuit is 
able to detect and mitigate Prime+Probe, Flush+Reload, and
Flush+Flush attacks on OpenSSL cryptography algorithms with 
an F-score of 1, and also detect and mitigate degradation 
of service on a vision application suite with an F-score of 0.9375.
The scheme poses low overheads: up to a maximum of 6\% in a no-attack scenario,
and up to a maximum of 11\% in an attack scenario. Biscuit also
reduced degradation of service in an attack scenario by up to 40\%
for one benchmark.
At the heart of the scheme is the generation and use of a cache-miss model,
which is inserted by the compiler at the entrances of
loop nests to predict the underlying cache misses. These beacons convey
the information to the scheduler, which uses them to 
co-schedule processes such that their combined cache footprint does not
exceed the maximum capacity of the last-level cache.
The scheduled processes are monitored for cache misses, and when an anomaly
is detected, the scheduler performs a search to isolate
the attacker. In addition, the scheme is able to detect the attacks quite early
and with high detection efficiency. 
We believe that due to the high precision, low overhead, and ability to deal with multi-tenancy, such a scheme is practicable.

%-------------------------------------------------------------------------------
% \section*{Acknowledgments}
%-------------------------------------------------------------------------------

%-------------------------------------------------------------------------------
\bibliographystyle{plain}
\bibliography{bibs/bibliography.bib, bibs/attack.bib, bibs/type.bib, bibs/techniques.bib, bibs/static.bib}

% \clearpage

\section{Appendix}

%%%%%%%%%%%%%%%%%%%%%%%%%%%%%%%%%%%%%%%%%%%%%%%%%%%%%%%%%%%%%%%%%%%%%%%%%%%%%%%%
\end{document}